\documentstyle[twocolumn,aps]{revtex} 
\begin{document} 
\draft
\title{ Structure of
 Eigenstates and Local Spectral  Density of States: \\
A Three-Orbital Schematic Shell Model}

\author{Wen-ge Wang$^{1,2}$, 
F.M. Izrailev$^{1,3,\dag}$, 
 G. Casati$^{1,4}$} 
\address{ $^1$ International Center for the Study of 
Dynamical Systems, University of Milan at Como, 
via Lucini 3, 22100 Como, Italy 
\\ $^2$ Department of Physics, Nanjing University, Nanjing 
210093, China 
\\ $^3$ Budker Institute of Nuclear Physics, 630090 
Novosibirsk, Russia
\\ $^4$ Istituto Nazionale di Fisica della Materia, Unit\'{a} 
di Milano, and I.N.F.N., 
Via Celoria 16, 20133, Milano, Italy} 
 
\twocolumn[
\maketitle
\widetext
%\vspace*{-1.0truecm}
\begin{abstract} 
\begin{center}
\parbox{14cm}{

The average shape of the Spectral Local Density of States (LDOS) and
eigenfunctions (EFs) has been studied numerically for a conservative
dynamical model (three-orbital Lipkin-Meshkov-Glick 
 model) which can exhibit 
strong chaos in the classical limit. The attention is paid to the 
comparison of the shape of LDOS with that known for random matrix models,
as well as to the shape of the EFs, for different values of the
perturbation strength. The  classical
counterparts of the LDOS has also been studied and 
found in a remarkable agreement with the quantum 
calculations. 
 Finally, by making use of a generalization 
of Brillouin-Wigner perturbation expansion, the form of long tails of 
LDOS and EFs is given  analytically and confirmed numerically.} 
\end{center}
\end{abstract}
 
\pacs{
\hspace{1.9cm} 
PACS number 05.45.+b}

]
\narrowtext

\section{Introduction}

  Recently, growing attention has been payed to the structure of the 
so-called Local Spectral Density of States (LDOS) in application to both 
disordered and dynamical systems which exhibit strong chaotic 
properties (see, for example, \cite{FGGK94,FBZ96,FCIC96,CCGI96,JSS96,MF97}). 
This quantity, known in nuclear physics as the ``strength function'', is of 
special interest since it gives information about the ``decay'' of a
specific unperturbed state into other states due to interaction. In particular,
the width of the strength function defines the effective ``life-time'' of the 
unperturbed basis state.

Typically, the shape of the LDOS is assumed to be of  Lorentzian
form (i.e.,  the ``Breit-Wigner shape''), as can be analytically
derived for  sufficiently weak coupling. However, in a direct computation
of Ce atom \cite{FGGK94} it was found that at relatively large distances 
from its center, the LDOS has an abrupt decay which is extremely fast (even 
faster than the exponential). This fact, which is quite  generic, 
 is due to the 
finite range of the 
interaction in the unperturbed energy basis \cite{C85,FLP89,GFG95}. 
As a result, matrix elements 
of a Hamiltonian describing a 
 realistic physical system, decay very fast 
away from the principal diagonal, thus leading to an effective band-like 
structure. 

Such  band structure of  Hamiltonian conservative systems can 
be compared to the one known for unitary evolution operators describing 
one-dimensional dynamical systems under 
 periodic perturbations, like the 
paradigmatic Kicked Rotator Model (KRM) \cite{CCIF79,I90}. Another example is
an ensemble of Hermitian Band Random Matrices (BRM), which  is used to 
describe quasi-1D disordered models in solid state physics (see, for example, 
\cite{FM94,I95} and references therein). The theory of such ``standard'' 
BRM is now well developed, see review \cite{FM94}; however, it can not 
be applied, verbatim, 
 to conservative systems like  isolated atoms, nuclei, 
atomic clusters, etc. The reason is that the Hamiltonians 
of these latter 
systems expressed in the basis of the reordered 
 unperturbed states, have 
 an additional leading diagonal corresponding 
to the energy density of the unperturbed Hamiltonian. Band random 
matrices with such an additional leading diagonal are known as Wigner 
Band Random Matrices (WBRM) (see \cite{CCGI96,FLP89,W55,FLW91,LF93,CCGI93}). 
Unlike the standard BRM, the theory of  WBRM is not well developed.  
On the other hand,  these matrices are currently 
 under close attention since they are believed to provide  an 
adequate description for complex systems 
(atoms, nuclei, clusters, etc), as well as
 for dynamical conservative systems with few 
degrees of freedom, which are chaotic in the classical limit. 

  In this paper we consider a specific dynamical model of this 
type, namely, the so-called Lipkin-Meshkov-Glick model \cite{LMG65}.  
 In our study
we follow the approach developed in \cite{CCGI96} where the structure of 
 LDOS and eigenfunctions (EFs) has been numerically investigated in 
detail for the WBRM. The main result of \cite{CCGI96} which stems from 
direct comparison of LDOS and EFs, is the discovery of the so-called 
''localization in the energy shell'' for conservative systems with chaotic 
behavior. It is of great interest to apply the approach suggested in 
\cite{CCGI96} to dynamical systems of interacting particles. 

    In this connection it may be interesting to remark that 
 it is possible to 
 relate specific properties of chaotic eigenstates to 
such observables as the occupation numbers for single-particle levels and
transition amplitudes ( see
details in \cite{FI97,FGI96,FIC96,FI97a}). 
The above approach \cite{FI97,FGI96,FIC96,FI97a} 
has been developed for the model of two-body random
interaction, by assuming completely random two-body  matrix elements.
Thus, it is important to extend this approach to dynamical 
systems of interacting particles with a chaotic dynamics.  

    The paper has the following structure. In Section II we describe the
three-orbital Lipkin-Meshkov-Glick model and discuss its general
properties. The classical limit is considered in Section III, where 
transition to chaos is studied in dependence on the strength of the 
perturbation. Section IV is devoted to the discussion of general 
properties of eigenstates and spectrum statistics for the quantum model. 
In Section V we numerically investigate the structure of  LDOS and 
eigenfunctions for different values of model parameters. In section VI, 
we present some analytical and numerical results for long tails of  
LDOS and eigenfunctions. Concluding remarks are given in Section VII.

\section{Three-orbital LMG model}

    The three-orbital Lipkin-Meshkov-Glick model \cite{LMG65}, or in 
short, the LMG model,   is known as some 
 simplification of the 
shell-model of the nucleus.  
It  was introduced also to  
check the validity of approximate many-body techniques, including 
the random-phase approximation and Bardeen-Cooper-Schrieffer
(BCS) theory.
The symmetric states of the LMG model, which we 
will use in our calculations, 
correspond to collective motions 
which may mimic the collective motion of the nucleus. 
  
  The model has $\Omega $ particles distributed among three 
single-particle orbitals with the same parity and angular 
momentum. Each  orbital is $\Omega $-fold degenerate. 
The ground, first and second excited orbitals are labeled by 
$r=0,1,2$, and 
the degenerate states within each orbital are labeled by 
 $\gamma =1,2,3, \cdots $,$\Omega $. 
 The energy of each orbital is denoted by $\epsilon _r$.
In our calculations, for simplicity (without the loss of generality), 
 we will set $\epsilon _0 =0$.

    The Hamiltonian of the model is 
\begin{equation} H=H^0+ \lambda V \end{equation}
where 
\begin{equation} \begin{array}{l}
\displaystyle H^0= \epsilon_1 
[ \sum_{\gamma =1}^{\Omega} a^{\dag }_{1\gamma}a_{1 \gamma} ]
+ \epsilon_2
[ \sum_{\gamma =1}^{\Omega} a^{\dag }_{2\gamma}a_{2 \gamma} ]
\\ \displaystyle V=\mu_1 
[ \sum_{\gamma =1}^{\Omega} 
 \sum_{\gamma '=1}^{\Omega}( a^{\dag }_{1\gamma}a_{0 \gamma} 
 a^{\dag }_{1\gamma '}a_{0 \gamma '} 
 + a^{\dag }_{0\gamma}a_{1 \gamma} 
 a^{\dag }_{0\gamma '}a_{1 \gamma '})] 
\\ \displaystyle \ \ \ +\mu_2[ \sum_{\gamma =1}^{\Omega} 
 \sum_{\gamma '=1}^{\Omega}( a^{\dag }_{2\gamma}a_{0 \gamma} 
 a^{\dag }_{2\gamma '}a_{0 \gamma '} 
 + a^{\dag }_{0\gamma}a_{2 \gamma} 
 a^{\dag }_{0\gamma '}a_{2 \gamma '})] 
\\ \displaystyle \ \ \ +\mu_3[ \sum_{\gamma =1}^{\Omega} 
 \sum_{\gamma '=1}^{\Omega}( a^{\dag }_{2\gamma}a_{1 \gamma} 
 a^{\dag }_{2\gamma '}a_{0 \gamma '} 
 + a^{\dag }_{0\gamma}a_{2 \gamma} 
 a^{\dag }_{1\gamma '}a_{2 \gamma '})] 
\\ \displaystyle \ \ \ +\mu_4[ \sum_{\gamma =1}^{\Omega} 
 \sum_{\gamma '=1}^{\Omega}( a^{\dag }_{1\gamma}a_{2 \gamma} 
 a^{\dag }_{1\gamma '}a_{0 \gamma '} 
 + a^{\dag }_{0\gamma}a_{1 \gamma} 
 a^{\dag }_{2\gamma '}a_{1 \gamma '})] 
\end{array} \label{HHH} \end{equation}
where $a^{\dag }_{r \gamma }$ and $a_{r \gamma}$ are fermionic
creation and annihilation operators obeying the usual anti-commutation 
relations,  and parameters $\lambda , \mu_1 , \mu_2, \mu_3, \mu_4$ describe 
the strength of the perturbation. 

 Hamiltonian  (\ref{HHH}) can be expressed in a much simpler 
form. To this end we introduce the  two-fermion 
operators 
\begin{equation} K_{rs} = \sum_{\gamma =1}^{\Omega} a^{\dag }_{r\gamma}
a_{s\gamma }  \ \  \ \ \ \ \ \ 
r,s=0,1,2. \label{Kij} \end{equation}
The operators $K_{00}, K_{11}$ and $K_{22}$ are   number operators 
of the orbitals 0, 1 and 2, and  
$K_{rs}$ for $r \ne s$  are particle raising 
and lowering operators, respectively.
The commutation relations for $K_{rs}$ are 
\begin{equation} [K_{rs},K_{r's'}]= K_{rs'}\delta_{r's} 
- K_{r's} \delta_{rs'}. 
\label{KK} \end{equation}
As a result, the Hamiltonian can be written as 
 a function of $K_{rs}$, 
\begin{equation} \begin{array}{l} 
H= H^0+ {\lambda } V 
\\ H^0=\epsilon_1 K_{11} + \epsilon_2 K_{22}
\\ \displaystyle V= \sum_{t=1}^4 \mu_t V^{(t)} 
\end{array} \label{H0} \end{equation} 
where
\begin{equation} \begin{array}{l} 
 V^{(1)}= K_{10}K_{10}+K_{01}K_{01}
 \\ V^{(2)}= K_{20}K_{20}+K_{02}K_{02}
\\  V^{(3)}= K_{21}K_{20} + K_{02}K_{12}
\\ V^{(4)}= K_{12}K_{10}+ K_{01}K_{21}.
\end{array} \label{V} \end{equation}

    The nine operators $K_{rs}$ have a very important property,  
namely, they are invariant under  interchange 
of the single-particle-state
 labels $\gamma$. Thus, the Hamiltonian  also is invariant 
under  interchange of $\gamma$ and  
conserves the permutation symmetry of the  labels $\gamma$. This 
makes  possible to divide  the Hilbert 
space into subspaces according to 
permutation symmetry. In our quantum calculations, we use a  
 subspace  composed by the so-called symmetric states. 
  A  convenient basis $|mn>$ 
for such subspace 
  can be obtained by operating the symmetric raising operators $K_{10}$ 
and $K_{20}$
 on  the  state  with all the 
$\Omega $ particles in the ground orbital, labeled by $|00>$, 
\begin{equation}  |mn> = C(m,n) K^m_{10} K^n_{20} |00>,
\label{mn} \end{equation}
where $C(m,n)$ is the normalizing coefficient. 
These states $|mn>$ are eigenstates of the number operators 
$K_{11}$ and $K_{22}$ with 
 $m$ being  the number of particles in the orbital 1
 and $n$  the number of particles in the orbital 2. 
By conservation of particles number, 
$\Omega = K_{00} + K_{11} + K_{22},$ 
 the population of the ground orbital is $\Omega -m-n$. 
 The dimension of the symmetric  subspace is 
$N=(\Omega +1)(\Omega +2)/2$.  
Notice that $|00>=a^{\dag}_{0\Omega} \cdots a^{\dag}_{02} 
a^{\dag}_{01}|00 \cdots 0>$,  and therefore,  $|mn>$ are antisymmetric 
 under  interchange of  label $\gamma$. 

    From Eq. (\ref{KK}) it is seen that the particle raising operators 
$K_{10}$ and $K_{20}$  commute 
(as a consequence of  their   two-fermionic 
feature), and therefore, the state $|mn>$ 
in (\ref{mn}) is symmetric under the interchange of 
the order of raising particles. In this sense, raising fermions 
from the ground orbital to the two excited orbitals is similar to 
creation of bosons. 
Indeed, resorting to the generator coordinate method approach 
to the dynamic group representation, a boson representation 
has been found for the symmetric states 
 \cite{XGWYF95,XWsjY87}.
 The above basis states  in the boson representation are 
\begin{equation} |mn> \ = \ 
\frac{(b_1^{\dag})^m (b_2^{\dag})^n}{\sqrt{(m+1)!(n+1)!}} 
|00>  \end{equation}
where $b^{\dag}_r$ are  creation 
 operators of bosons. 
 The relations between 
the operators  $K_{rs}$ and the creation 
and annihilation operators of bosons $b^{\dag}_r$ and 
$b_r$ are 
\begin{equation} \begin{array}{l}
K_{rs} = b^{\dag}_r b_s 
\\ K_{r0} = K^{\dag}_{0r} =  b^{\dag}_r \sqrt{\Omega -b^{\dag}_1b_1 
-b^{\dag}_2b_2} 
\\ {} [b_r,b^{\dag}_s]=\delta_{rs}, \ \ {  } [b_r,b_s]= 
[b^{\dag}_r, b^{\dag}_s]=0 
\end{array}
\label{Kb} \end{equation} 
for $r,s=1,2$.
Making use of these relations 
it is easy now to obtain the expressions for the matrix elements 
of $V^{(t)}$ 
\begin{equation} \begin{array}{l}
<m'n'|K_{10}K_{10}|mn> = \\  \sqrt{(\Omega -m-n)(\Omega -m-n-1)
(m+1)(m+2)} \delta_{m',m+2} \delta_{n',n} 
\\ <m'n'|K_{20}K_{20}|mn> = \\  \sqrt{(\Omega -m-n)(\Omega -m-n-1)
(n+1)(n+2)} \delta_{m',m} \delta_{n',n+2} 
\\ <m'n'|K_{21}K_{20}|mn> = \\  \sqrt{m(\Omega -m-n)
(n+1)(n+2)} \delta_{m',m-1} \delta_{n',n+2} 
\\ <m'n'|K_{12}K_{10}|mn> = \\  \sqrt{n(\Omega -m-n)
(m+1)(m+2)} \delta_{m',m+2} \delta_{n',n-1}. 
 \end{array} \label{Hele} \end{equation} 
which can also be found from the commutation relations 
(\ref{KK}), see \cite{MKZ88}. 

    The states $|mn>$ are  eigenstates of $H^0$ 
with  eigenenergies 
\begin{equation} E^0_{mn} = m\epsilon_1 + n\epsilon_2. 
\label{E0} \end{equation}
It is convenient to rearrange  the eigenstates 
of $H^0$ in  order of increasing energy,  
 and we  
  will   label them by $|\phi_i>$ 
\begin{equation} H^0 |\phi _i> = E^0_i |\phi_i>, 
\ \ \ \ \ \ \ \ E^0_{i+1} \ge E^0_i. 
\end{equation}
Correspondingly, the eigenstates of the total 
Hamiltonian $H$, also reordered in   energy, 
will be labeled by $|\psi_{\alpha}>$
\begin{equation} H |\psi_{\alpha}> 
=E_{\alpha}|\psi_{\alpha}>.
\label{Sch} \end{equation}

    In our numerical computations on the LMG model, we 
take $\Omega =40$; therefore, the dimension of the symmetric 
subspace is $N=861$. As to the choice of $\epsilon_1$ and 
$\epsilon_2$, we assume,  
 without  loss of generality,  $\epsilon_1=1.1$ 
and $\epsilon_2=1.61$. 

    The expressions for $V^{(t)}$ in Eqs.\ (\ref{V}) and (\ref{Hele}), 
show that each $V^{(t)}$ couples the basis state  
$|\phi_i>$  with only two other states $|\phi_{j}>$ with the same 
energy difference $d^{(t)}=|E^0_i- E^0_{j}|$ where  
$d^{(1)}=2\epsilon_1, \ d^{(2)} = 2\epsilon_2 , \ 
d^{(3)} = 2\epsilon_2 -\epsilon_1 , \ 
d^{(4)} = 2\epsilon_1 - \epsilon_2$. 
    The average coupling strength 
 $v^2=<V_{ij}^2>$ can be found by  
 averaging over  the non-zero matrix elements only. 
 Similarly, for each 
$V^{(t)}$ one can introduce $(v^{(t)})^2= 
<(V^{(t)}_{jk})^2>$ with the average taken over only 
the non-zero matrix elements of $V^{(t)}$, respectively. 
Therefore,  $\rho^{(t)} \equiv v^{(t)}/d^{(t)}$ 
is a natural measure of the strength of $V^{(t)}$ 
with respect to the energy distance between the basis states 
 coupled by $V^{(t)}$.
The parameters $\mu_t$ in Eq. 
(\ref{H0}) are determined by the condition that $\rho^{(t)}=1$
 for $t=1,2,3,4$, 
so that, on average, the relative 
strengths of $V^{(t)}$ 
 are the same. Under this condition, we have $\mu_1 \approx 0.0116$, 
$\mu_2 \approx 0.0169$, $\mu_3 \approx 0.0158$, $\mu_4 \approx 
0.00439$, and  the estimate of the average coupling strength 
 is $v \approx 2.24$. 

    The global structure of the Hamiltonian matrix is presented in 
Fig.\ 1, where  points represent  
     non-zero off-diagonal elements 
 $<\phi_i|V|\phi_j> $ of the 
Hamiltonian $H$.  
As one can see the matrix is sparse and band-like. More precisely,  
 the non-zero elements of the perturbation 
 form only eight curves since
a basis state $|\phi_i>$ is coupled by $V$ 
to at most eight   other basis states. 
The two inner curves result from the contribution of 
$V^{(4)}$, while  
the two outer curves come from 
$V^{(2)}$. Since $d^{(1)}=2.2 \approx d^{(3)}=2.12$, the curves  
corresponding to  $V^{(1)}$ and $V^{(3)}$ are very close to 
each other and are not separated in the figure.
The half-band width $b$
depends on $E^0_i$ and can be analytically estimated to be 
$b \approx 3E^0_i/ \epsilon_2$ for $E^0_i <E'$, and 
$b \approx 2\Omega -6(E^0_i-E')/ \epsilon_2$ for $E^0_i > E'$, 
where $E' \approx 2\Omega \epsilon_2 /3$. 
In particular, the maximum width (in the center of the band) is 
$b_{max} \approx 2\Omega  $.  

    The unperturbed density of states $\rho(E^0)$ of $H^0$ is shown 
in Fig.\ 2a, and  turns out to be in agreement with the estimate
$\rho(E^0) \approx b/ d^{(2)}$. The perturbed density of 
states $\rho(E)$  is shown in Fig.\ 2b for $\lambda =2.0$. 
For a better comparison of $\rho(E)$  with $\rho(E^0)$, 
it is convenient to rescale to the same total energy interval, 
namely $\rho(E) \to \rho_{\nu }(E) 
= \nu  
\rho(E \nu)$, where $\nu$ is 
\begin{equation} 
\nu =\frac{(E_{861}-E_1)}
{(E^0_{861}-E^0_1)}
\label{es} \end{equation}
Here, $E_{861}$ and $E^0_{861}$ are the highest eigenenergies 
of  $H$ and 
$H^0$, respectively. The ground state energy of the perturbed 
Hamiltonian has also been shifted to coincide with the unperturbed 
one. From the result of such  rescaling, shown in Fig.\ 2c, 
one can conclude that the rescaled perturbed density of states is similar 
to the unperturbed one. For weaker 
perturbations, $\lambda \le 1$, the correspondence is much better. 
From Fig.\ 2c one can also see that 
 the peak of $\rho(E)$ is shifted a little towards the 
center of the spectrum. Moreover, for even stronger perturbations, 
the peak has been found to be at the center of the spectrum. 
These properties of the global structure of the Hamiltonian and of  
the density of states will be used below when we discuss other  
properties of the model. 

\section{The Classical limit}

    The classical limit of the  symmetric subspace of the LMG model can 
be obtained by two methods. One method was used in 
\cite{MKZ88} and consists in the direct study of  the motion of 
coherent states in the limit  $\Omega \to \infty$. 
The other method, which will be used here, is based on the boson 
representation. However, since the boson representation of the 
symmetric states is obtained via the  
coherent state representation, the two methods are 
basically equivalent. 

    In order to obtain the classical limit, we introduce the 
transformation,  
\begin{equation} b^{\dag}_r = \sqrt{\frac{\Omega}2} 
(q_r -ip_r), 
 b_r = \sqrt{\frac{\Omega}2} 
(q_r +ip_r), 
\label{bpq} \end{equation}
for $r=1,2$. According to (\ref{Kb}) and (\ref{bpq}), $q_r$ and $p_s$ 
obey the following 
commutation rules, 
\begin{equation} [q_r, p_s]=\frac i{\Omega} \delta_{rs}. 
\end{equation}
Therefore, the factor $1/\Omega$ plays the role of Planck constant. 
Letting the particles number $\Omega \to \infty$ 
while keeping constant the following parameters 
\begin{equation} \begin{array}{l}  
\epsilon_1' = \epsilon_1 \Omega,  \ \ \ 
\epsilon_2' = \epsilon_2 \Omega, 
\\ \mu_t' = \mu_t \Omega^2, \ \ \ \ \ \ \ \ \ \ \ \ \ \ 
t=1,2,3,4, 
\end{array} \end{equation}
 one obtains the classical counterpart of
 the Hamiltonian $ H$ , 
\begin{equation} 
H_{cl} = H^0_{cl} + \lambda  V_{cl} \end{equation}
where 
\begin{equation} \begin{array}{l}
\displaystyle H^0_{cl} = \frac{\epsilon_1'}2 (p_1^2 +q_1^2) 
+ \frac{\epsilon_2'}2 (p_2^2 +q_2^2) 
 \\ \displaystyle V_{cl}=\sum_{t=1}^4 \mu_t'V_{cl}^{(t)}  
 \\ \displaystyle = \mu_1'(q_1^2-p_1^2) (1-G/2)
+\mu_2'(q_2^2-p_2^2) (1-G/2)
\\ \displaystyle 
\ \ + \frac{\mu_3'}{\sqrt{2}}[(q_2^2-p_2^2)q_1 + 
2q_2p_1p_2]\sqrt{1-G/2}
 \\ \displaystyle \ \  + \frac{\mu_4'}{\sqrt{2}}[(q_1^2-p_1^2)q_2 + 
2q_1p_1p_2]\sqrt{1-G/2}
\end{array} \label{Hc}  \end{equation} 
with $G=q_1^2 + p_1^2 + q_2^2 + p_2^2 =2(b^{\dag}_1b_1 + 
b^{\dag}_2b_2)/ \Omega \le 2$. 
Notice that
 the perturbation 
$V_{cl}$ depends also on  momentum variables. 

    In order to understand the qualitative properties of the classical 
model,  we have plotted  the Poincare surfaces of sections 
at different energies. 
 As in \cite{XGWYF95,MKZ88}, it was found that 
regular regions of phase space 
 are gradually destroyed when $\lambda $ increases. 
However, due to  the 
specific form of the classical Hamiltonian 
 $H_{cl}$ in Eq.\ (\ref{Hc}), 
 it has been found that the motion on low and high 
energy surfaces  can exhibit more chaotic features than  on the medium 
energy surfaces. Three typical examples for $\lambda =0.9$ are 
shown in Fig.\ 3. The first figure, 
Fig.\ 3(a), shows the surface of section at 
energy   $E=10$. 
 It can be seen  that 
 trajectories on this energy surface are 
chaotic except in a small region. The next 
Fig.\ 3(b) corresponds to  the energy   $E=39$, in the 
middle of the energy region. Here one can distinguish
 three regular islands. 
Finally, Fig.\ 3(c) corresponds to  the energy  $E=57$ in the high 
energy region. Here most trajectories are seen to be chaotic
with some remaining regular islands. Note that 
the central region in Fig.\ 3(c) 
is energetically inaccessible.  
 When $\lambda$ increases to 2.0, it has been 
found that almost all  regular islands disappear and 
the system is nearly totally chaotic (Fig. 4). 

    The above peculiar behavior is due to 
 the particular structure  of the perturbation 
 $V_{cl}$. Indeed,  
at energy $E=39$, the value of $G$ 
 can be closer to 2 than for the case with $E=10$. 
 Therefore, due to the terms 
containing $(1-G/2)$ in (\ref{Hc}), the  perturbation 
at the energy $E=10$ is stronger than at $E=39$. 
 For high energies, instead, 
since $G$ is quite close 
to 2  the derivatives $\partial V_{cl}/ \partial p_i$ and
 $\partial V_{cl} / 
\partial q_i$ can be  large and, as a consequence, the 
 motion at high energies  ($ E=57$) is more irregular 
than that at the middle ones.

\section{General properties of Eigenstates and Spectrum  Statistics}

    In the above section, we have discussed   
the classical counterpart of the LMG model. In particular, 
we showed that for not very large perturbations ,  
 the classical 
 motion on low and high energy 
surfaces is more irregular than on the middle ones. 
In this section, we study some general properties of the 
quantum model, which are related to  
 the above  classical features. 

    In Figs.\ 5, 6 we show four typical eigenstates 
$|\psi_{\alpha}>$ of the total Hamiltonian $H$ for $\lambda =0.9$
 and $\alpha =50-53$ 
and 430-433, respectively, in the basis states $|\phi_i>$ 
(many-particle states of the unperturbed Hamiltonian $H^0$). 
 For low levels $\alpha =50-53$, 
the states $|\psi_{\alpha}>$  mainly occupy the 
region  $i=0 \div 200$ of the basis states $|\phi_i>$, 
and the expansion coefficients  look random 
 in the region. For the levels in the middle of the energy spectrum, 
 $\alpha =430-433$, the components of the eigenstates 
  $|\psi_{\alpha}>$ are 
mainly distributed  in the region $ i=200 \div 700$, 
but the expansion coefficients  
 do not appear as completely random.
For example, the coefficients 
 $<\phi_i|\psi_{\alpha}>$ for $\alpha$ =433  
 seem to be random, without any structure;  
instead, for $\alpha$ =431, they 
look sparse and some structure is seen. 
  These figures 
suggest that eigenstates with low energies 
are more chaotic  in the region  
$i \in [1,200]$  than those with  
middle energies in the region  
$i \in [200,700]$. This is also confirmed  by 
nearest-level-spacing 
distributions. In Fig.\ 7 we plot the nearest-level-spacing 
distributions $P(s)$ for eight different 
 regions in the  energy spectrum of 
$H$ for $\lambda =0.9$. 
In order to achieve better statistics we have diagonalized the 
Hamiltonian  with five different values of  $\lambda$ close to 
$\lambda =0.9$  
and  put together the  unfolded sequences $\Delta E_{\alpha}$.  
As expected, the histograms of $P(s)$ for the lowest and highest 
energy regions are closer to the Wigner-Dyson distribution (dashed lines) 
than to the Poisson distribution (dashed-dot lines), 
while  for $\alpha$ in the interval [440,550] $P(s)$ is 
closer to the Poisson distribution. On the other hand, 
when $\lambda$ increases to 2, the level spectrum distribution 
 $ P(s)$ become very close to 
the Wigner-Dyson distribution even  in the middle energy region.

The above numerical results are related to  
 properties of the perturbation
 $V$, which are determined by 
 the four operators 
 $K_{10}K_{10}$, $K_{20}K_{20}$, $K_{21}K_{20}$ 
and $K_{12}K_{10}$. So one needs to  study only 
non-zero matrix elements of these four operators. For example, 
according to Eq.\ (\ref{Hele}), for a fixed basis state 
$|\phi_i>$, there is only one basis state $|\phi_j>$ for which the 
matrix element  $<\phi_j|K_{10}K_{10}|\phi_i>$
 is non-zero. Therefore, the 
non-zero matrix elements of $K_{10}K_{10}$ can be regarded as 
a function of $i$ only. The same is true for the other three operators. 
    The dependence of  non-zero matrix elements of the above 
four operators on $i$, 
 is presented in Fig.\ 8. Several 
features can be seen from this figure. Firstly, on average, the 
non-zero matrix elements of $\mu_1 K_{10}K_{10}$  are relatively  
large in the low energy region. Secondly, apart from  the two edges, 
the average values of the non-zero matrix elements 
 of $\mu_2 K_{20}K_{20}$  are similar in 
different energy regions. However, in the middle of the energy 
region, the operator $\mu_2 K_{20}K_{20}$ has many 
very small non-zero matrix elements.
 Thirdly,
the matrix elements of $\mu_3 K_{21}K_{20}$ 
 are relatively large,  
on average, in the high energy region. 
 Finally, the variation of the 
matrix elements of $\mu_4 K_{12}K_{10}$  in different 
energy regions is not so large as compared to the other three 
operators. As a result,  the perturbation 
is stronger in the low and high energy regions than  in the middle 
 energy region. 

%    To see the relation between the above properties of the quantum 
%model and  those of its classical counterpart discussed in the 
%previous section, one should note the following relations 
%between the quantum operators $b_r^{\dag}, b_r$ and the classical 
%variables $q_r,p_r$. That is, in the limit  $\Omega \to 
%\infty$, 
%\begin{equation} \begin{array}{l} \displaystyle 
%b_r^{\dag} b_r \longrightarrow \frac{\Omega}2 (q_r^2+p_r^2), 
%\ \ \ \ \ \ \ r=1,2
%\\ \displaystyle \sqrt{1- \frac{b_1^{\dag}b_1 + b_2^{\dag}b_2}
%{\Omega}} \longrightarrow \sqrt{1 - \frac G2}. 
%\end{array} \label{bpq} \end{equation}
%According to Eq.\ (\ref{Kb}), the influence of $\sqrt{\Omega - 
%b_1^{\dag}b_1 -b_2^{\dag}b_2}$ on the properties of the quantum 
%system is similar to that of $\sqrt{1-G/2}$ on its classical 
%counterpart. 

\section{Structure of LDOS and Eigenfunctions}

    In this section we discuss the shape of  LDOS and of eigenfunctions 
for the LMG model, we study 
the classical counterpart of the LDOS, and finally
 we discuss to what extent the LMG model can be associated 
with a band random matrix model. 

\subsection{ Structure of LDOS and eigenfunctions}

    The so-called  local spectral density of states (LDOS) for an 
unperturbed state $|\phi_j>$  is 
defined as 
\begin{equation} 
w_j(E) = \sum_{\alpha} |C_{\alpha j}|^2 \delta (E-E_{\alpha})
\label{ldos} \end{equation}
where $E_{\alpha}$ is the eigenenergy of the perturbed eigenstate 
$|\psi_{\alpha}>$ and $C_{\alpha j} = <\phi_j|\psi_{\alpha}>$. 
The function $w_j(E)$, also known as the ``strength function'' 
or ``Green spectra'', is quite  important for the understanding 
of generic properties of the quantum model. In particular, the LDOS shows how 
the unperturbed state $|\phi_j>$ is coupled to the exact states 
$|\psi_{\alpha}>$ with the specific energy $E_{\alpha}$. 
The width of this function (``spreading width'') defines the 
energy range associated with the ``life time'' of an unperturbed 
state $|\phi_j>$. 

    The form of the LDOS for band random matrices has been 
analytically studied  by Wigner \cite{W55}, see also \cite{FCIC96}. 
Particularly, it was shown that when  perturbation is not large, 
 the LDOS has the form
\begin{equation} 
w_{BW}(E-E^0_j) = \frac{\Gamma /2 \pi }{(E-E^0_j)^2 + 
\Gamma ^2/4}  
\label{BW}  \end{equation}
which is nowadays known as the Breit-Wigner (BW) law. Here, $\Gamma$ 
is the half-width of the distribution.  
For larger perturbations, the form of 
LDOS becomes model dependent and in the intermediate region can 
be approximately 
 described 
by a Gaussian distribution \cite{CFI97}. 

    Another important quantity is the shape of eigenfunctions 
(EFs)
\begin{equation} 
W_{\alpha}(E^0 ) = \sum_{j} |C_{\alpha j}|^2 \delta (E^0-E^0_j)
\label{ef} \end{equation}
 in the unperturbed energy basis.   
    In our numerical calculations of the LDOS and EFs for the LMG model, 
in order to suppress fluctuations, we have taken averages 
over 200 of individual distributions in the interval 
$331 \le j \le 530$  for the LDOS and 
$331 \le \alpha \le 530 $  for the EFs. 
The averaged distributions will be denoted  by 
$w(E)$ and $W(E^0)$, respectively. 
Before averaging, we should  first express 
$w_j(E)$ and $W_{\alpha}(E^0)$ with respect to their centroids, 
respectively. For the LDOS, 
the centroid of $w_j(E)$ is just $E^0_j$ \cite{FBZ96}
\begin{equation}
E_j^0 = \sum_{\alpha} E_{\alpha} |C_{\alpha j}|^2,  
\end{equation}
so that we can express the LDOS as $w_j(E-E_j^0)$. On the other hand,
the centroid of $W_{\alpha}(E^0 )$, labeled by $e_{\alpha}$, is 
defined by 
\begin{equation}
e_{\alpha} = \sum_j E_j^0 |C_{\alpha j}|^2 ,
\label{ea} \end{equation}
and $W_{\alpha}$ can be expressed as a function of the shift
 $(E^0-e_{\alpha})$.

    The dependence of  the shape of  LDOS and EFs  
 on the perturbation  is presented in 
Fig.\ 9. The left column gives the LDOS and the right column 
shows the EFs. (Notice that the vertical 
scale depends on the value of 
 $\lambda $.). The first remark  
 is that 
 the shapes of  LDOS and 
of EFs are quite similar when the perturbation is not large
 ($\lambda \le 0.9$). 
 On the other hand, 
with increasing  $\lambda $, they start to 
 deviate from each other. 
Another result is that for not large perturbation,  
 $\lambda \le 0.9$, there seem 
to be large peaks  which 
 are not washed out by the averaging 
 process over 200 distributions. 
 In fact, they come from dynamical interference 
(correlation) effects, 
 which  will be explained 
 in the appendix B. 
 
 The dashed curves in Fig.\ 9 
correspond to the best fit to the BW   and 
the dashed-dot curves correspond to the best fit to the Gaussian.  
 The fitting was made here for the central parts of the LDOS and 
EFs. Specifically,  neglecting the long tails,  
 we chose for the fitting only the data with $w$ and $W$ larger 
than 0.01.

    In Fig.\ 9 it can be seen that
for  $\lambda 
=0.5$  the central parts of both distributions 
 can be well fitted  by both the BW and 
Gaussian forms (however, the agreement with the Gaussian 
extends to the region of the tails).  
 For  stronger perturbation  $\lambda =0.9$,  
the distributions
 can still be fitted quite well by the Gaussian 
form. Finally, when  the perturbation is very strong, for example 
$\lambda =2.0$, the LDOS and EFs deviate, as expected,  
from both the BW and the Gaussian form. 

    The difference between the LDOS and the EFs for large 
perturbations $\lambda =2.0$,  is quite evident.  
 However, one should note that 
 the LDOS is plotted in the perturbed energy basis 
while the EF in the unperturbed one. Therefore, in order to make 
the comparison meaningful, one  should rescale the distribution 
in a proper way. 
We used the same rescaling  as in Fig.\ 2c. 
After this rescaling (see Fig.\ 10) they look
 more similar to each other  than in Fig.\ 9.

\subsection{The classical limit of LDOS}

 The classical 
counterpart of LDOS, in short, the classical LDOS,  labeled by
 $w_{cl}(E-E^0_j)$, can be defined as 
 the probability that a phase point, which belongs to the 
 torus corresponding to the  quantum 
numbers $m_j$ and $n_j$ of $|\phi _j>$, 
 has total energy $E$ \cite{CCGI96}.
It is  expressed 
as a function of the distance $(E-E^0_j)$ 
where $E^0_j$ is the unperturbed energy  of the torus $m_j, n_j$. 
According to Eq.\ 
(\ref{bpq}), in the limit  $\Omega \to \infty$ we have 
\begin{equation}
\frac{b^{\dag}_rb_r}{\Omega } = \frac{ (p_r^2 + q_r^2)}2 
\end{equation}
Thus, the torus corresponding to $m_j$ and $n_j$ is that with 
$(p_1^2 +q_1^2) =2m_j/ \Omega$ and 
$(p_2^2 +q_2^2) =2n_j/ \Omega$.

In analogy to  the quantum case,  
 the classical LDOS was averaged over
 200  different tori. 
   In Fig.\ 11, we show a comparison between the quantum and  classical LDOS 
 for $\lambda =0.3,0.5,0.9$ and 2.0. 
As discussed in appendix B, the LDOS for $\lambda =0.3$ depends on 
 strong dynamical quantum correlations; as it is seen from the figure, 
 in the classical limit these correlations disappear.  
 At the same time, the shape 
of the classical LDOS is, on average, close to the quantum one 
apart from the tails which are of quantum origin and are due to 
tunneling effects. 
The data also show that with  increasing the perturbation, the 
agreement becomes better. 
The main difference 
 for $\lambda =2.0$ is 
that in the center the LDOS is lower than the classical LDOS. This may 
be related to a result given in Ref. \cite{FCIC96} that 
there is a local minimum in the center of the LDOS for relatively  
strong perturbations. 

%     Now let us compare the shapes of the four classical 
%LDOS in Fig.\ 11. They look quite similar apart from different 
%scales. The detailed analysis shows that  indeed they obey the 
%scaling law 
%\begin{equation} w_{cl}(E-E^0) = \lambda '
%w_{cl}' [(E-E^0) \lambda '] {}
%\label{cws} \end{equation}
%where $w_{cl}(E-E^0)$ is the  classical 
%LDOS for $\lambda =1$. 
% To prove 
%the relation  (\ref{cws}), let us 
% consider a set of phase points on the energy 
%surface  $H^0_{cl}=E^0$. 
% Denote now by $p(v)$ the distribution of the values of the perturbation
% $V_{cl}$,  
%and by $p'(v)=p(v/ \lambda ' )/ \lambda'$ 
%   the distribution of the values of $\lambda 'V_{cl}$
%The distribution of $E=E^0+V_{cl}$ and $E'= 
%E^0+ \lambda 'V_{cl}$ are now $p(v+E^0)$ and $p(v/ \lambda '+E^0)
%/ \lambda '$, respectively. 
%Then the distributions of $E-E^0$ and $E'-E^0$ of the above phase 
%points are just $p(v)$ and $p'(v)=p(v/ \lambda' )/ \lambda'$,  
%which leads to  Eq.\ 
%(\ref{cws}). 

\subsection{A band random matrix model}

Recently it has been  shown that  realistic 
conservative systems with chaotic properties can be approximated by
  band random matrices 
\cite{FGGK94,C85,FLP89,GFG95}. Therefore, it is  
 very interesting to check whether the 
Hamiltonian of the LMG model, which does not contain 
any random matrix element, 
 see Eq.\ (\ref{Hele}), can be associated with an ensemble of 
random matrices.  
To be as close as possible to the dynamical model, 
we introduce here band random matrices of the form 
\begin{equation} H_{ran} = H_0 +\lambda V_{ran}
\label{Hran} \end{equation} 
where $H_0$ is the same as for the LMG model, see Eq. (\ref{H0}),  
and $V_{ran}$ is obtained by replacing the non-zero 
matrix elements of $V$ of the LMG model by random numbers 
with Gaussian distribution. The mean value of the  
matrix elements $(V_{ran})_{kl}$  is zero and 
the variance, 
averaged over the non-zero matrix elements, is taken to be the same as 
 in the dynamical LMG model $<(V_{ran})^2_{kl}>=<V_{kl}^2>$. 

    Numerical data for both 
    the LDOS and EFs of $H_{ran}$ are presented in Fig.\ 12. 
Averages have also been taken over 200 LDOS and EFs, respectively, 
in the central region of the spectrum, as in the calculations of 
the LMG model. 
Interestingly, these results are similar to those found for 
Wigner band random matrices \cite{FCIC96}, that is, for
small perturbations the central part of the LDOS is of the BW
 form, while  
in  the transition region when the perturbation is relatively strong,  it 
can be fitted to the Gaussian form; for stronger perturbations  
it can be fitted approximately to the semicircle law 
\begin{equation} w(E)= \frac 2{\pi R^2_0} \sqrt{R^2_0 -E^2}. 
\label{sem} \end{equation}
    
    From  Fig.\ 12 and Fig.\ 9, it can be seen that the shape 
of  LDOS and  EFs of $H_{ran}$ for $\lambda =0.3, 0.5$ and 
$0.9$ are much more smooth than the corresponding LDOS and EFs of 
 the LMG model. This can be explained by the randomness of 
$(V_{ran})_{kl}$, see  appendix B. 
Another feature  is that 
the central parts of the LDOS and of the EFs  
for $\lambda =0.3 \div 0.9$ in Fig. 12 are lower than those for the 
Lipkin model. This also is an effect of interference. Finally, 
comparing the two figures, it can be seen that for $\lambda 
\le 0.9$ the central parts of  LDOS and  EFs of the LMG model are 
roughly similar to those of $H_{ran}$. Thus, when perturbation is 
not very large, the LMG model can be associated with the above 
band random matrix model (\ref{Hran}).

\section{ Long Tails of LDOS and Eigenfunctions}

    In the previous section, we have discussed the central 
parts of  LDOS and  EFs. In this section we study their 
long tails with the help of perturbation theory since, 
as  discussed in appendix A, long tails are always 
in the perturbative region. 

First we consider  the case of  small $\lambda $  for which 
the coefficients $|C_{\alpha j}|$ 
  decrease very fast as $|E_{\alpha} 
-E^0_j|$ increases.   
Let us start with the left tails of EFs, $E^0_j << E_{\alpha}$, 
 with $E_{\alpha}$ in the middle  energy region. 
From Eq. (\ref{GBW}), we have 
\begin{equation} C_{\alpha j}^2 \equiv |<\phi_j| 
\psi_{\alpha}>|^2 = | \frac{<\phi_j|\lambda V|\psi_{\alpha}>} 
{E_{\alpha}-E^0_j}|^2 
\label{C2aj} \end{equation}
As indicated in section II, 
there are only eight basis states, 
that can be coupled  with a given basis state $|\phi_j>$ by 
the perturbation $ V$.
Denote these states by $|\phi _l>$, $l=l_1, l_2, 
\cdots l_8$ in order of increasing energy. 
Notice that since the energy differences 
$d^{(t)}$ are generally much larger than the local 
level spacings (about 0.1, on average, see Fig. 2a),
 there are many 
basis states located between each two of the above 
 eight states $|\phi_{l}>$. Therefore,  
if the term  $|C_{\alpha k}|^2$ decreases fast enough with decreasing 
 $k$ in the region $E^0_k << E_{\alpha}$,
 the component $|C_{\alpha k}|$ for $k=l_8$ will be much larger than 
the sum of the other seven components, then, 
\begin{equation} 
 C_{\alpha j}^2 \approx 
 \frac{(\lambda V_{jl8}^{(2)})^2} 
{(E_{\alpha}-E^0_j)^2 } C_{\alpha l8}^2 
\label{lct} \end{equation}
    Following the  
procedure  given in App. D of Ref. \cite{FGGK94}, one obtains the following  
 estimate for  the left tails of the averaged EFs  
\begin{equation} W(\xi ) \propto exp \{ -2 \xi 
ln (\frac{\xi}{e} \frac{2\epsilon_2}{\lambda v^{(2)}}) \} 
\label{left} \end{equation}
 where $\xi =|E_{\alpha} - E^0_j|/2\epsilon_2$ and 
$v^{(2)}$ is the average strength of $V^{(2)}$ in the low energy region 
($V^{(2)}_{jl8}$ in Eq. (\ref{lct}) is in the low energy region). 

For the right tails of EFs,  $E^0_j >> E_{\alpha}$,similar arguments 
 lead again to Eq. (\ref{left}) 
 with  $v^{(2)}$ changed to the 
average strength of $V^{(2)}$ in the high energy region, which, 
according to Fig.\ 
8, is approximately equal to the one in the low energy region. 

    To obtain an expression for the tails of  LDOS for $|\phi _j>$ 
with $E_j^0$ in the middle energy region,  
we assume that  in the regions of tails the  
shapes of different eigenstates are similar on average, that is 
\begin{equation} 
\overline {C^2_{\alpha l8}} \approx 
\overline{C^2_{\alpha'j}} 
\label{cc} \end{equation}
 for $E_{\alpha'} -E^0_j \approx E_{\alpha}-E^0_{l8}$. 
Then Eq. (\ref{lct}) gives 
\begin{equation} 
 C_{\alpha j}^2 \approx 
 \frac{(\lambda V_{jl8}^{(2)})^2} 
{(E_{\alpha}-E^0_j)^2 } C_{\alpha ' j}^2 
\label{lct2} \end{equation}
and, for the two tails of LDOS, 
 one obtains the same expression as Eq. (\ref{left})  
with $\xi =|E_{\alpha}- E^0_j|/2\epsilon_2$   
and $v^{(2)}$  the average strength of $V^{(2)}$ in the middle 
energy region 
($V^{(2)}_{jl8}$ in Eq. (\ref{lct2}) is in the middle energy region). 

    When $\lambda$ is not small, as indicated in appendix B 
more research work is needed in order to obtain
an analytical expression for the tails of EFs and LDOS. 
 However, we can assume that the tails obey a law somewhat similar to 
(\ref{left}) with $v^{(2)}$ changed to $v$ (since when $\lambda $ 
is not small,  
 the tails are  determined not only by $V^{(2)}$, but also 
by the other $V^{(t)}$). Here similar to the small $\lambda $ case, 
 $v$ is the average strength 
of the perturbation in the corresponding regions: for the left 
and right tails of  EFs, the average 
should be taken in the low and high energy regions, respectively, 
while for the two tails of  LDOS the average should be taken in the 
middle energy region. According to Fig. 8,  
the value of $v$ in the low energy region  
 is larger than that in the high energy region, so 
 the right tail of the EF should drop faster than the left one, 
while the two tails of the LDOS should be similar. 

    We have numerically computed 
    the tails of both the LDOS and  EFs for the case of  weak 
perturbation $\lambda =0.1$ and the results 
 are shown in Fig.\ 13 in  logarithm scale. 
Each point represents an average over 200 states. 
 It can be seen that  
the tails of LDOS and EFs 
 are quite close  to each other in agreement with the fact 
that, for small $\lambda $, they  
 obey the same law given by Eq. 
(\ref{left}) with similar values of $v^{(2)}$. 
Also the agreement between the numerical 
results and the analytical prediction is quite good. 

    When the perturbation increases to  
 $\lambda $=0.3, the 
 tails  begin to deviate from 
the prediction  (\ref{left}).  
 However, it has been found that for $\lambda \ge
0.3$ the tails  can be fitted quite well to the expression 
\ (\ref{left}) with $v$ instead of  $v^{(2)}$  and $\xi$ given by 
\begin{equation}
\xi = (\frac{|E_{\alpha}-E^0_j| -x_0}{10 \epsilon_2 })^2
\label{xi2} \end{equation}
where $x_0$ is an adjusting parameter. 
    As an example 
 we present in Fig.\ 14 the results for $\lambda =0.9$.  
 The LDOS and its fitting curve with $x_0=13$ 
are given in Fig.\ 14a. It can be seen that the agreement is 
quite good not only in the long tail regions, but also in the 
regions quite close to the central part. Fig.\ 14b gives the 
tails for the EF and the fitting curve with $x_0=16$. 
The figure also verifies the prediction given  above 
that the right tail drops faster than the left one for the EF, 
while the two tails are similar for the LDOS.

    For the band random matrix model 
 (\ref{Hran}), as indicated in appendix B, the long tails of EFs and 
LDOS obey the same law given by Eq. (\ref{left}). 
 This has been confirmed 
by numerical results. As an example, in Fig.\ 15, we present 
the tails of LDOS for $\lambda =0.9$ and the prediction 
given by Eq. (\ref{left}).
 The agreement between  numerical data and 
the analytical result is again quite well in the long tail regions.

\section{Conclusions and Discussions}

  In this paper we have studied the Lipkin-Meshkov-Glick (LMG) model in the 
many-particle basis of non-interacting particles. The main attention has 
been paid to the structure of the local spectral density of states (LDOS)
in comparison with that of eigenfunctions (EF) defined in the unperturbed 
energy basis.  

  Due to its dynamical nature, the properties of the LMG model strongly 
depend on the energy region. Namely, for strong enough perturbation 
$\lambda $, chaotic properties of the model for low and high excitation 
energies are stronger than in the middle of the energy spectrum. This fact 
is explained both on the ground of peculiarities of the quantum model 
and of its classical counterpart. In particular, the eigenstates in the 
middle of the spectrum are more regular as compared with the eigenstates 
for low and high energies; correspondingly, the level spacing distribution
is close to the Wigner-Dyson type at the edges of the spectrum, unlike 
in the center of the spectrum where deviations from the Wigner-Dyson have 
been detected. 

  One of the main questions addressed in our study is the dependence of the
shape of LDOS and EFs on the perturbation strength. Numerical analysis has 
shown that for relatively weak perturbations the form of LDOS is close to 
the Breit-Wigner apart from  tails. This fact is in accordance with  
several observations for models with random interaction described by different
random matrix ensembles \cite{FGGK94,FCIC96}. However, the detailed study 
of LDOS and EFs for  the LMG 
 model in the region of not very strong interaction, 
have revealed remarkable correlations which can be analytically explained. 
In fact, these correlations are due to the dynamical nature of the model 
and they are found to be washed out for stronger interaction. 

  With the  increase of the perturbation, the form of the LDOS changes and for 
a quite moderate perturbation it is well described by a Gaussian. This
observation is quite interesting in view of recent numerical data for complex
atoms \cite{FGGK94} and heavy nuclei \cite{FBZ96}, where the form of the 
LDOS was found to be quite close to a Gaussian. The same effect (the change
of the form of  LDOS from the Breit-Wigner to the Gaussian-like) has 
also been found in numerical investigations of Wigner Band Random Matrices 
\cite{CFI97}. Therefore, our data for the dynamical LMG model indicate that
the above fact is of quite generic nature and occurs also in dynamical models 
exhibiting strong chaos in the classical limit. Finally, when the perturbation 
is very strong, the LDOS has been found to have a quite specific form which 
is related to the peculiarity of the model under consideration.

 Another problem is the relation between the shape of the LDOS and that of EFs. 
We have found that after a proper rescaling, the shape of the EFs is similar 
to the shape of the LDOS, if the perturbation is not very strong. 
 This result indicates  the
absence of localization in the energy shell which has been found in Wigner
Band Random Matrices \cite{CCGI96}. 

 As was noticed in  \cite{CCGI96}, the shapes of both LDOS and EFs
have an analogy in the classical limit. In our paper the relation between 
 LDOS and the corresponding classical quantity has been checked for the 
first time in a dynamical model with chaotic classical counterpart (see 
also recent paper \cite{BGIC97}). Numerical analysis of the classical model 
has shown that the form of LDOS is close to its classical counterpart
if the perturbation is not very weak. This fact allows to expect that
(in semiclassical regions) the 
global structure of both the LDOS and EFs can
be directly found from the corresponding classical model. This is 
important in view of recent results \cite{FI97,FI97a} revealing the direct
connection of the shape of EFs to the partition function of isolated systems.

 Of special interest is the question about the applicability of random
matrix models to a given dynamical system in a chaotic region. There are 
many examples when full random matrices or band random matrices can serve
as good models for the description of statistical properties of spectra and 
eigenstates of dynamical models. However, in these examples the two-body 
nature of an interaction is, typically,  not taken into account.
In this paper we have
carefully analyzed the possibility of a random matrix description of the model
in the energy region where the corresponding classical system can be 
treated as a chaotic one. Specifically, we have used the same unperturbed part
of the Hamiltonian, but with off-diagonal matrix elements chosen at random with 
the same variance as in the original dynamical model, keeping zero matrix 
elements which are due to the 
 specific form of the interaction. This approach 
also allows us to reveal to what extent the underlying correlations of the 
dynamical model are essential for its statistical description. Numerical 
results with such a random model have shown a quite good agreement for 
global properties of the LDOS and chaotic EFs. In particular, the shapes 
of  LDOS and EFs turn out to remain similar in a large range of the 
perturbation strength. On the other hand, the form of LDOS and EFs in the 
random model turns out to be more smooth and do not reveal any regular 
deviations due to quantum dynamical correlations, which, for a weak 
interaction, are significant in the dynamical model.

 Finally, the question of long tails of  LDOS and EFs has been studied in
detail, both analytically and numerically. For this, a generalized
approach has been developed based on the standard Brillouin-Wigner
perturbation expansion. Namely, the perturbation theory has been extended
for strong perturbation in the region of long tails. This has allowed to find
the analytical form for the long tails of LDOS and EFs.  Numerical data 
have shown a good agreement with the analytical predictions.  

\section{Acknowledgments}

    The authors are grateful to I. Guarneri for valuable discussions. 
W.-G. W. and F.M.I. wish to thank their colleagues of the University of 
Milano at Como for the hospitality during the period when this work was 
done. F.M.I. acknowledges the support received from
 the Cariplo Foundation for Scientific Research as well
as the partial support by the Grant No. INTAS-94-2058.
Partial support by the National Basic Research Project ``Nonlinear 
Science'', China, Natural Science Foundation of China and 
Post-doctoral Foundation of China is also gratefully 
acknowledged by W.-G.W.

\appendix 
\section{  Brillouin-Wigner 
perturbation expansion of eigenstates }

Here we  introduce a generalization of the so-called 
Brillouin-Wigner perturbation expansion \cite{Z69,XWwgY92}, 
which can be shown to be valid even for strong perturbations. 
In particular, we show that 
long tails  are always in the 
perturbative region and this explains also some previous 
results \cite{FGGK94,FCIC96,CCGI96}. 
    
    First, we divide the 
set of basis vectors, i.e., the eigenstates 
of $H^0$, into two parts, $\{|\phi_i>, \ i=p_1, \cdots 
p_2 \}$ and $\{ |\phi_i>, \ i=1, \cdots p_1-1,p_2+1, 
\cdots N \}$.
Correspondingly, there are two projection operators 
\begin{equation} P=\sum_{i=p_1}^{p_2} |\phi_i><\phi_i|, 
\ \ \ \ \ Q=1-P. 
\label{PQ} \end{equation}
The subspaces related to $P$ and $Q$ will be called 
in the following the {\it P} and 
{\it Q subspaces }, respectively.
We now split an arbitrary eigenstate $|\psi_{\alpha}>$
of $H$ into two orthogonal parts
\begin{equation} |\psi_{\alpha}> = 
|t> + |f>, \ \ \ \ \ \ <t|f>=0
\end{equation}
where $|t>=P|\psi_{\alpha}>$ and $|f> = Q |\psi_{\alpha}>$.
Applying $Q$ to the Schr\"{o}dinger equation (\ref{Sch}), 
we readily obtain 
\begin{equation} [ E_{\alpha}-H^0]|f> = 
Q\lambda V|\psi_{\alpha}>, 
\end{equation} 
then, 
\begin{equation} 
|\psi_{\alpha}>= |t> + \frac 1{E_{\alpha} - H^0} 
Q\lambda V|\psi_{\alpha}>,  
\label{GBW} \end{equation}
in which the eigenstate $|\psi_{\alpha}>$  can be 
either normalized or not. 
We remark that in the particular case in which $|t>$ is chosen as 
a single basis state $|\phi_i> \ (i=\alpha)$ and $E_{\alpha}$ taken as 
\begin{equation} E_{\alpha} = E_i^0 + <\phi_i|\lambda V 
|\psi_{\alpha}>, 
\end{equation}
then the iterative expansion of  Eq.\ (\ref{GBW}) is just the 
Brillouin-Wigner perturbation expansion, in which the 
eigenstate $|\psi_{\alpha}>$ is not normalized.  

    Since our purpose  is to study the structure of  
eigenfunctions and of LDOS, and not to solve the eigenvalue  
problem, we take $E_{\alpha}$ in Eq.\ (\ref{GBW}) as a constant, 
i.e., its exact (unknown) value which can be find out by other methods.  
 Then the iterative expansion  of Eq. (\ref{GBW}) gives  
\begin{equation} \begin{array}{l}
\displaystyle 
|\psi_{\alpha}> = |t> + \frac 1{E_{\alpha}-H^0}
Q\lambda V|t> 
\\ \displaystyle 
\ \ \ + \frac 1{E_{\alpha}-H^0}Q\lambda V
  \frac 1{E_{\alpha}-H^0}Q\lambda V|t> + \cdots
\\ \displaystyle 
\ \ \ + (\frac 1{E_{\alpha}-H^0}Q\lambda V)^{n-1}|t>
\\ \displaystyle 
\ \ \ + (\frac 1{E_{\alpha}-H^0}Q\lambda V)^n|\psi_{\alpha}>. 
\end{array} \label{EGBW} \end{equation}
If the last term on the right hand side of Eq.\ (\ref{EGBW}), 
denoted by $T_n$, tends to zero when $n \to \infty$, 
one has a {\it generalization  of Brillouin-Wigner perturbation 
expansion (GBWPE)}, which gives an exact expression for  
 $|\psi_{\alpha}>$ in terms of $|t>, \ E_{\alpha},\  \lambda V$ 
and $H^0$:  
\begin{equation} \begin{array}{l}
\displaystyle 
|\psi_{\alpha}> = |t> +
\frac 1{E_{\alpha}-H^0}Q\lambda V|t>
\\ \displaystyle  
\ \ \ \ + (\frac 1{E_{\alpha}-H^0}Q\lambda V)^2|t>+ \cdots
\\ \displaystyle  
\ \ \ \ + (\frac 1{E_{\alpha}-H^0}Q\lambda V)^q|t> + \cdots
\end{array}  \label{e-psi} \end{equation}
 For example, for a Hamiltonian 
matrix with band structure and  perturbation $ V$ coupling 
 a basis state to at 
most $b$  other states, if the $Q$ subspace is such chosen  that 
for $|\phi_j>$ in the $Q$ subspaces 
 $|E_{\alpha} - E^0_j| \ge b V_{max}$, 
where $V_{max}$ is the maximum of the absolute value of $(\lambda 
V_{ij})$,  
then $T_n$ will approach 
zero when $n$ goes to  infinity.

    For each state $|\psi_{\alpha}>$, there exists 
a $P$ subspace with the minimum number of basis states (denoted
 by $P_{min}$) 
 and  correspondingly 
a $Q$ subspace with the maximal number
 of basis states ( denoted by $Q_{max}$)
 that ensure  Eq.\ (\ref{e-psi}) 
to hold. Clearly, from Eq.\ (\ref{e-psi}), the components of the state 
$|\psi_{\alpha}>$ in the $Q$ subspace are in the perturbative region. 
Therefore,  the expansion of the exact state $|\psi_{\alpha}>$ in
the basis $|\phi_j>$ 
is naturally divided into two parts: the non-perturbative 
part $|t_{min}> \equiv P_{min}|\psi_{\alpha}>$  and the 
perturbative part $|f_{max}> \equiv Q_{max}|\psi_{\alpha}>$. 
Tails of $|\psi_{\alpha}>$ 
 are always in the perturbative region and can 
be studied with the GBWPE. The same approach can be applied to 
the LDOS, which can also be divided into two parts, perturbative 
and non-perturbative.

\section{A ``path'' approach to  perturbative expansion}  

    In order to  study the perturbative part   of 
 $|\psi_{\alpha }>$ in GBWPE given in Eq. (\ref{e-psi}),  
we make use of the concept of path, in analogy to that in  the 
  Feynman's path integral theory
\cite{F65}. 
 First we discuss the case of   small   
perturbations. In this case, the term $|t>$ can 
be chosen as the basis state $|\phi_i>$ with the smallest 
value of $|E_{\alpha} -E^0_i|$. For an arbitrary $|\phi_j>$ 
in the Q subspace, Eq.\ (\ref{e-psi}) gives 
\begin{equation} \begin{array}{l} 
\displaystyle 
C_{\alpha j} \equiv <\phi_j|\psi_{\alpha}>  
\\ \displaystyle
\ \ \ \  = \frac {\lambda V_{ji}}{E_{\alpha} - E^0_j} 
+ \sum_{k \in Q} \frac {\lambda V_{jk}}{E_{\alpha} - E^0_j} 
\frac {\lambda V_{ki}}{E_{\alpha} - E^0_k}
\\ \displaystyle
\ \ \ \ + \sum_{k,l \in Q} \frac {\lambda V_{jk}}{E_{\alpha} - E^0_j} 
\frac {\lambda V_{kl}}{E_{\alpha} - E^0_k} 
\frac {\lambda V_{li}}{E_{\alpha} - E^0_l} \cdots 
\end{array} \label{cja} \end{equation} 
 We now  term a sequence  $j \to k_1 \to 
\cdots \to k_{q-1} \to i$ 
{\it a path of q paces} from $j$ to $i$,
 if the matrix elements of the perturbation 
 $V_{kk'}$ corresponding to  each pace is non-zero. 
Attributing a  factor $\lambda V_{kk'}/(E_{\alpha} -E^0_k)$ to each 
pace $k \to k'$, 
 the contribution of a path  to $C_{\alpha j}$  
is the product of  the  factors of all its  paces.  
  Then, the $q$-th term of $C_{\alpha j}$ on the right hand side 
of Eq. (\ref{cja}) can be rewritten as 
\begin{equation}
F_q(j \to i) = \sum_s f_{q,s}(j \to i)
\label{Ff} \end{equation}
where $s$ indicates the  paths with $q$ paces from $j$ to $i$ and 
 $f_{q,s}(j \to i)$ is the contribution of the path 
$s$  to $C_{\alpha j}$. Defining $A(j \to i)$ as 
\begin{equation} 
A(j \to i) = \sum_{q=q_0}^{\infty} F_q(j \to i), 
\label{AF} \end{equation}
where $q_0$ is the number of the paces of the shortest path from 
$j$ to $i$,  we have 
\begin{equation}
C_{\alpha j} = A(j \to i) \label{CAS} \end{equation}
for  small $\lambda V$. 

    For the general case, the term  $|t>$ cannot be chosen 
as a single  basis vector, but must be a
 superposition of some basis vectors. Writing 
\begin{equation} |t> = \sum_{i=p_1}^{p_2} t_i |\phi_i>,  
\end{equation}
$C_{\alpha j} =<\phi_j|\psi_{\alpha}>$ can be expressed as 
\begin{equation} \displaystyle  
C_{\alpha j} = \sum_{i=p_1}^{p_2}A(j \to i)\cdot t_{i} 
\label{CA} \end{equation}

    For Hamiltonians with band structure, paths have 
qualitatively different features for small and large 
perturbations. When the perturbation is small, 
the difference  $(p_2-
p_1)$, see Eq. (\ref{PQ}), of the $P_{min}$ subspace is smaller than the 
band width $b$. A path which start from $j < p_1$, can reach 
points $k$ larger than $p_2$, in other words, a path 
can ``cross'' the $P$ region.  
 Thus, the difference $E_{\alpha}- E_k^0$ can be 
both positive and negative. This is important 
for the parameters we have chosen in section II, which make all the 
non-zero matrix 
elements of V  positive. When $\lambda $ is so large 
that $p_2-p_1$ is larger than $b$, paths starting 
from $j<p_1$ can no longer reach $k>p_2$, i.e., can 
not ``cross'' the $P$ region.  
 Hence the values of  $E_{\alpha}-E_k^0$ are either always 
positive or negative for a path. 

    The above concept of path  is very useful in studying 
Hamiltonians with band and sparse structures. For the LMG 
model it is especially useful since the perturbation 
$V$  couples a basis state  
to at most eight  others as shown in Fig.\ 1. 
In fact, we have a simple method to 
find  out all possible paths from $j$ to $i$. First we should write 
the basis vectors $|\phi_i>$ 
 in their equivalent forms $|m_i,n_i>$ with $m_i$ being the number 
of particles in the orbital 1 and $n_i$ the number of particles 
in the orbital 2.  
 Let us denote the changes of $m$ and $n$ along a path 
by $\Delta m=m_i-m_j$ and $\Delta n=n_i-n_j$, 
and the changes of $m$ and $n$ for a pace $r$ from $k=k_l$ to $k'=k_{l+1}$ 
by $\delta m_r=m_{k'}-m_{k}$ and $\delta n_r=n_{k'}-n_{k}$. Then, if 
the path has $q$ paces, we can write  
\begin{equation} \begin{array}{c} \displaystyle
\Delta m = \sum_{r=1}^q \delta m_r
\\ \displaystyle \Delta n = \sum_{r=1}^q \delta n_r. 
\end{array} \label{Dmn} \end{equation}
Since  the interchange of $r$ does not influence the sum, Eq. (\ref{Dmn}) 
is also satisfied by 
some other related paths. 
According to Eq.\ (\ref{Hele}), there are only eight possible 
pairs of $\delta m$ and $\delta n$
\begin{equation} \begin{array}{l}
\delta m= \pm 2, \ \delta n=0 \ \ \ \ \ \ for \ V^{(1)}
\\ \delta m= 0, \ \delta n=\pm 2 \ \ \ \ \ \ for \ V^{(2)}
\\ \delta m=\mp 1, \ \delta n=\pm 2 \ \ \ \ \ \ for \ V^{(3)}
\\ \delta m=\pm 2, \ \delta n=\mp 1 \ \ \ \ \ \ for \ V^{(4)}, 
\end{array} \label{dmn} \end{equation}
therefore, in order to find out all possible $q$-pace
 paths from $j$ to $i$,
 one should  just  find out all possible 
combinations of $q$ pairs of the possible $\delta m$ 
and $\delta n$ in Eq.\ (\ref{dmn}) that satisfy Eq.\ 
(\ref{Dmn}) (if all the intermediate points 
are in the $Q$ subspace).

When $\lambda $ is small, the term  $|t>$ 
in Eq. (\ref{e-psi}) can be chosen as one basis vector $|\phi_i>$, 
and the expanding coefficient $C_{\alpha j}= <\phi_j|\psi_{\alpha}>$ is 
$A(j \to i) =\sum _q F_q$, see Eq.\ (\ref{CAS}). 
 As an example, let us consider the $C_{\alpha j}$ for which 
 $j$ is determined by 
$\Delta m =m_i-m_j=2, \ \Delta n =n_i -n_j=0$, i.e., the two 
unperturbed  states  $|\phi_i>$ 
and $|\phi_j>$ are coupled directly by the term $K_{10}K_{10}$ 
in Eq.\ (\ref{V}). Denote the 
number of the $q$-pace paths 
 with  positive contributions 
 $f_{q,s}$ in $F_q$ (see Eq. (\ref{Ff}))  by 
$N_{q+}$ and the number of paths with  negative $f_{q,s}$ by $N_{q-}$. For 
 the shortest path with one pace,  
$N_{1+}=1, N_{1-}=0$. There is no path with 2 paces. For $q=3$, 
there are 12 paths (e.g., $K_{02}K_{02} \to K_{20}K_{20} \to 
K_{10}K_{10}$), $N_{3+}=11, N_{3-}=1$, and for  $q=4$, we have $N_{4+}=11, 
N_{4-}=7$. Clearly, the number of short paths with positive $f_{q,s}$ 
is much larger than that of short paths with negative $f_{q,s}$. 
Therefore, $|C_{\alpha j}|$ is quite large and 
 the dynamical interference effect is  strong.  
 Similarly, for the other seven $|\phi_j>$ that 
can be coupled directly to $|\phi_i>$ by the other terms of $V$, 
 $|C_{\alpha j}|$ are also quite large. On the other hand, for 
$|\phi _j>$ which are not coupled directly to $|\phi _i>$ 
by $V$, $|C_{\alpha j}|$ are smaller. 
 In fact, the two peaks beside the main 
peak in fig.\ 9 for $\lambda =0.3$ come from $C_{\alpha j}$ for the eight 
 states $|\phi_j>$ coupled directly by $V$ to the state $|\phi_i>$. 
 
    For the band random matrix model discussed in section V, 
 since the non-zero matrix elements of $V_{ran}$ 
have random signs, we have $N_{q+} \approx N_{q-}$, and the dynamical 
interference effect should  be much smaller than that in the LMG model.

     Finally, let us give some discussion about long tails of LDOS 
and EFs when $\lambda $ is not small. In this case, 
 we should use Eqs. (\ref{AF}) and (\ref{CA}). 
 In the region of long tails, 
for each $A(j \to i)$ the 
main contribution comes from the term $F_{q_0}$. To see this, we give an 
estimate of $F_q(j \to i)$. Since one basevector can be coupled 
to at most eight  others by $V$, the number of possible 
paths with $q$ paces from $j$ to $i$ is $M^q$ where $0 < M < 8$. Then for long 
tails we have  
\begin{equation} 
F_q(j \to i) \approx (\frac{Mv}{E_{\alpha} - \widetilde E^0_j})^q. 
\end{equation}
where $v$ is the average coupling strength of the perturbation $V$ and 
for large $q$, $\widetilde E_j^0$ is approximately equal to $ E_j^0$. 
When $|E_{\alpha} -E^0_j|$ is large enough, the value $F_q(j \to i)$ 
 decreases very fast with  increasing  $q$ and the main 
term is the one with the shortest path ($q_0$ paces) from $j$ to $i$.
 Therefore, $C_{\alpha j}$
can be estimated as 
\begin{equation} 
C_{\alpha j} \approx \sum_i t_i F_{q_0}(j \to i) 
\label{CtF} \end{equation}

    For the LMG model, due to the dynamical 
 interference (correlation)  effects as discussed 
above, $F_{q_0}(j \to i)$ for different $i$ 
 may be equally important. In fact, for large $q_0$, although 
the number of paces is large, the number of paths is also large  
and some of the paces at $k$  may have relatively small
 values of $|E_{\alpha} - 
E^0_k|$. Thus, it is difficult to obtain an analytical expression for 
the tails from Eq. (\ref{CtF}).

    For the band random matrix model 
 (\ref{Hran}), 
the signs of the non-zero matrix 
elements of $V_{ran}$ are random. Therefore, as discussed above, 
  dynamical  interference effects  are not so large as 
in the LMG model,  and the largest term on the right hand side 
of Eq. (\ref{CtF}) is the one with the smallest number of paces $q_0$. 
Since $d^{(2)}$
 is the largest among the four $d^{(t)}$, 
 the shortest path from $j$ to $i$ 
 mainly consists of paces resulting from $V^{(2)}$. Then Eq. (\ref{lct}) 
also holds and the same expressions for the long tails of  EFs and 
 LDOS as in  (\ref{left})  can be found.

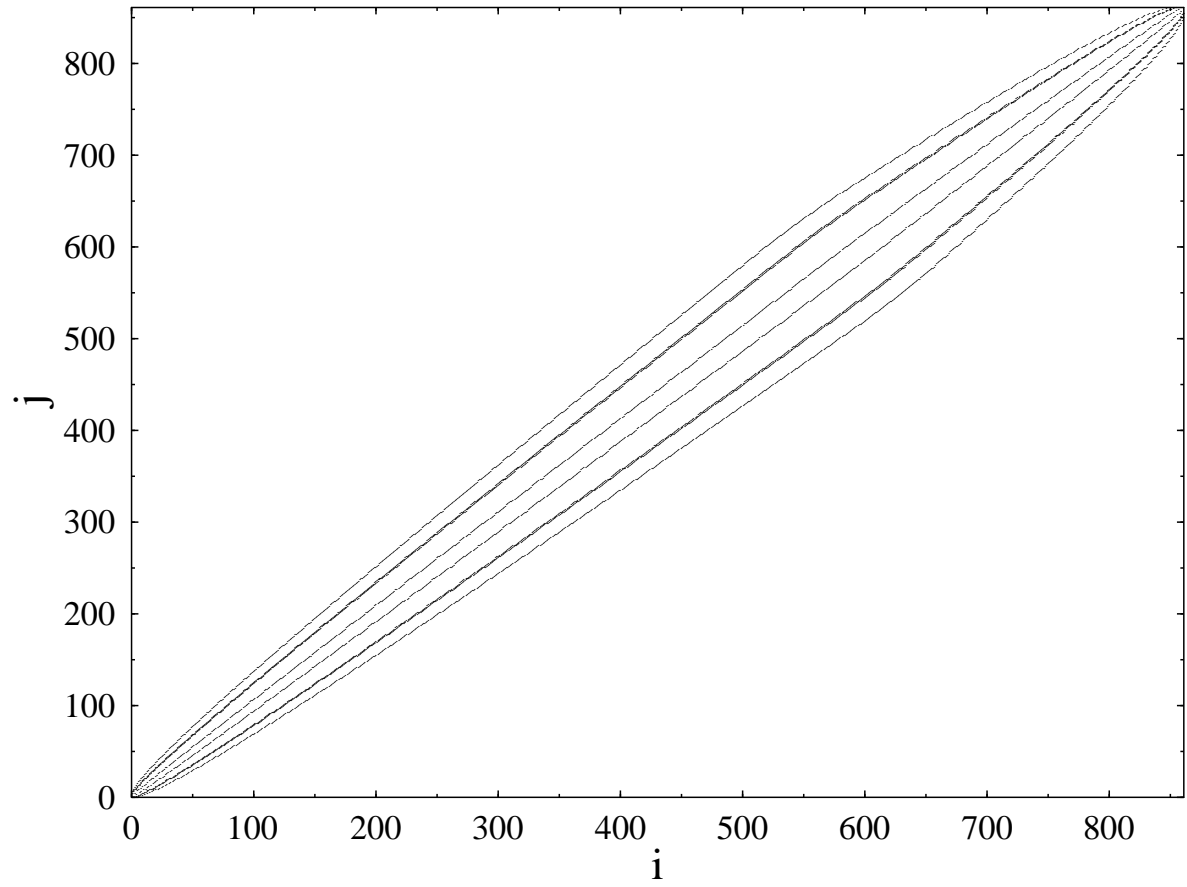
\begin{figure}\caption{Global structure of the Hamiltonian matrix. 
The points represent the non-zero off-diagonal elements 
 $<\phi_j|V|\phi_i> $ of the 
Hamiltonian $H$. } 
\end{figure} 

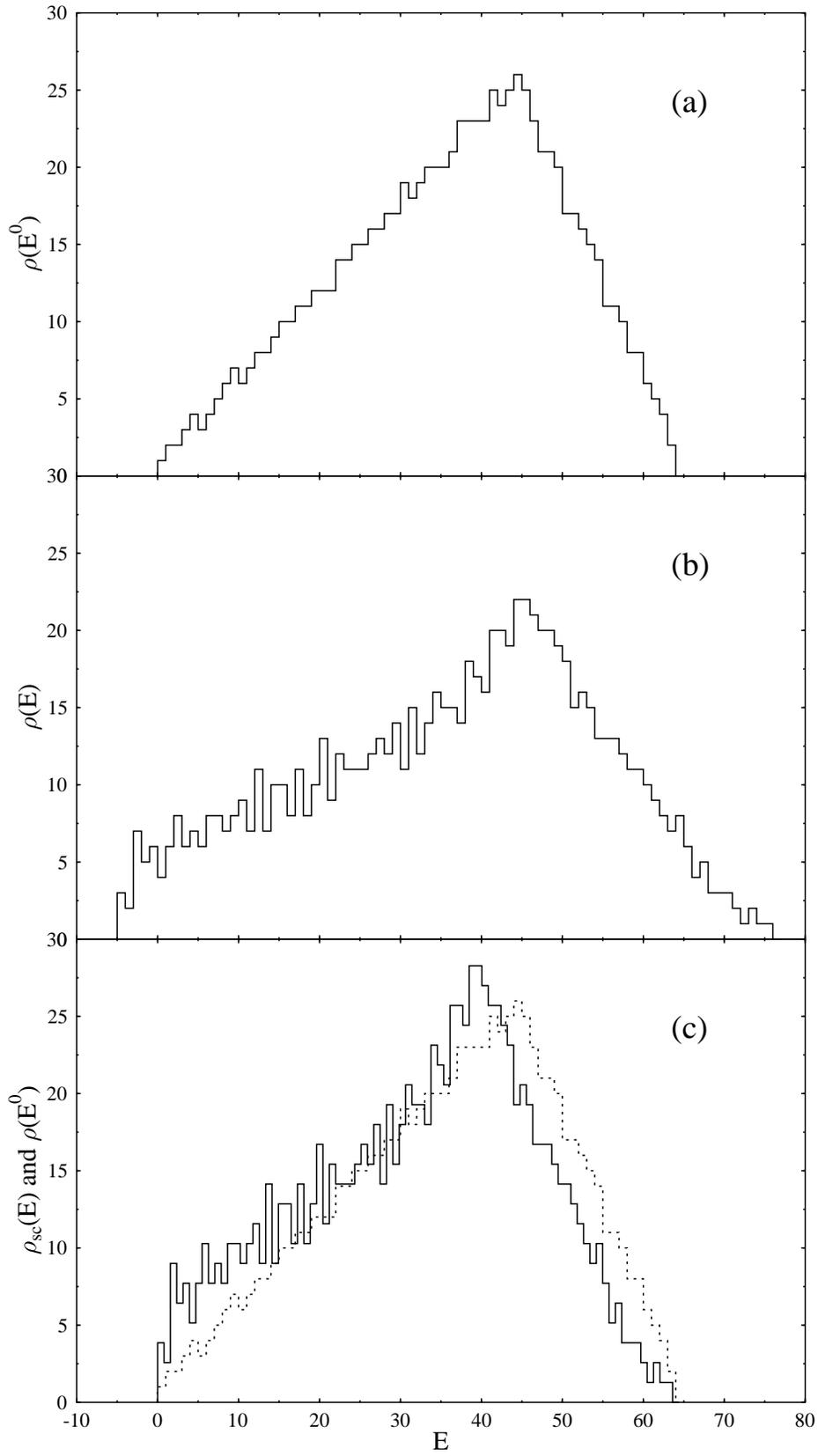
\begin{figure} \caption{ (a) Unperturbed density of states 
 $\rho(E^0)$ of  
 $H^0$.  (b) Density of states $\rho(E)$ of the 
 system $H$ for $\lambda =2$. (c) The rescaled density of states 
 $\rho_{\nu }(E)$ 
for $\lambda =2$ (solid histogram) compared with the unperturbed 
density of states $\rho(E^0)$ (dashed histogram). 
} \end{figure} 

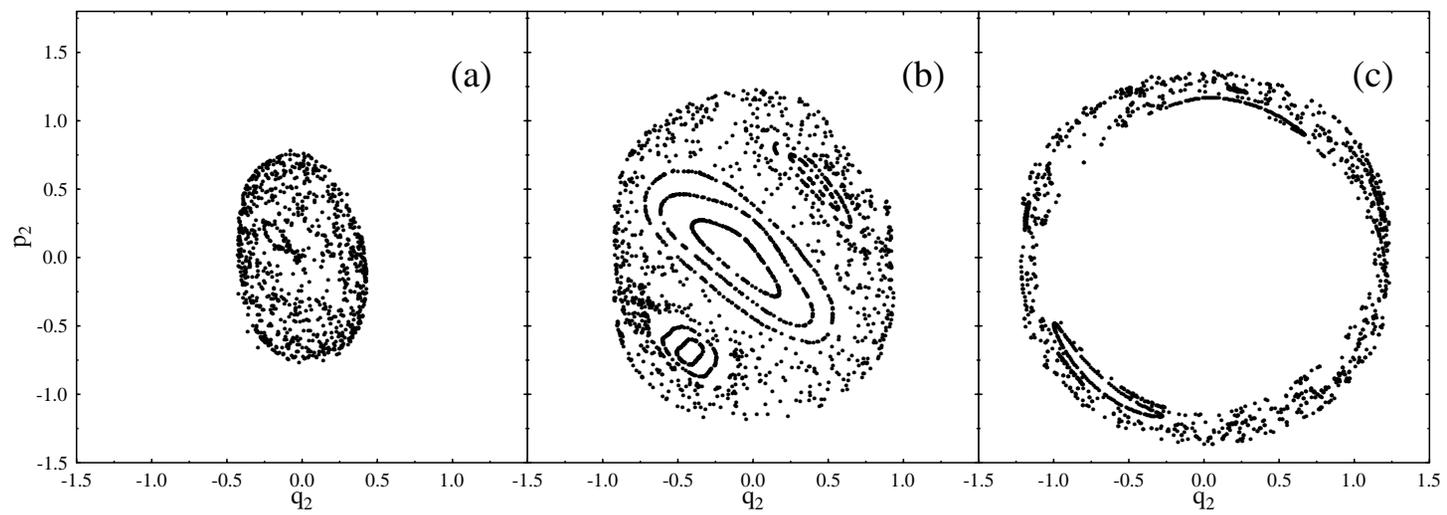
\begin{figure} \caption{ Poincare surface of 
 sections  on the 
$(q_2,p_2)$ plane with $q_1 =0$ for $\lambda =0.9$ and 
 (a) $E=10$, (b) $E=39$ and (c) $E=57$. 
} \end{figure} 

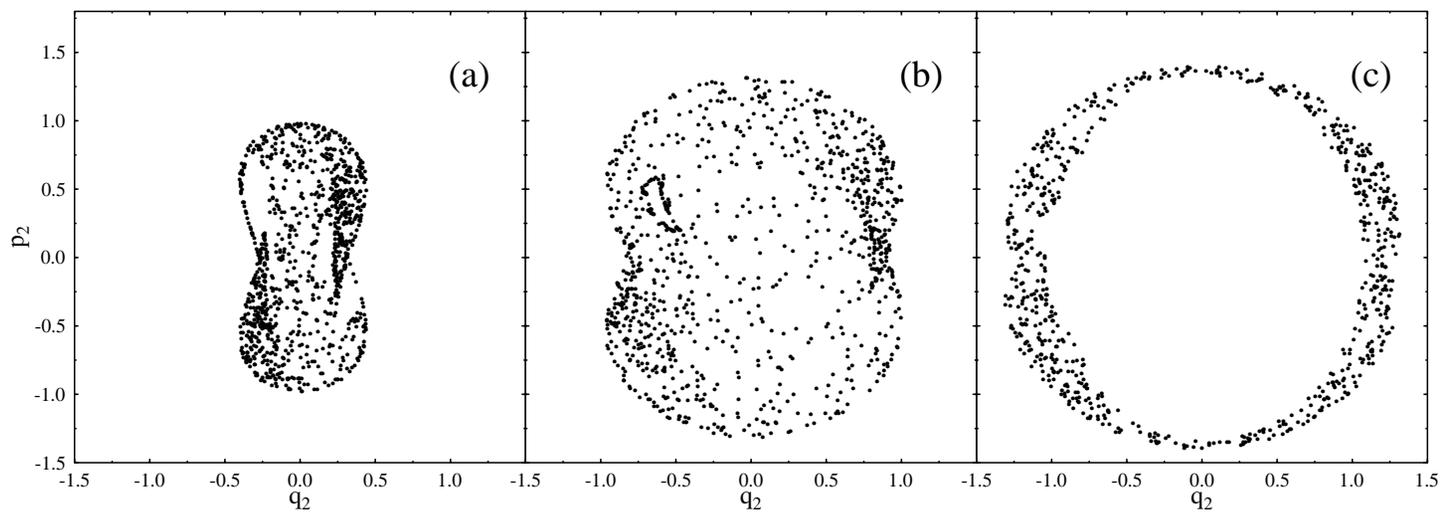
\begin{figure} \caption{ Same as in Fig.3 for $\lambda =2$ and 
 (a) $E$=4, (b) $E$=43 and (c) $E$=62. 
} \end{figure} 

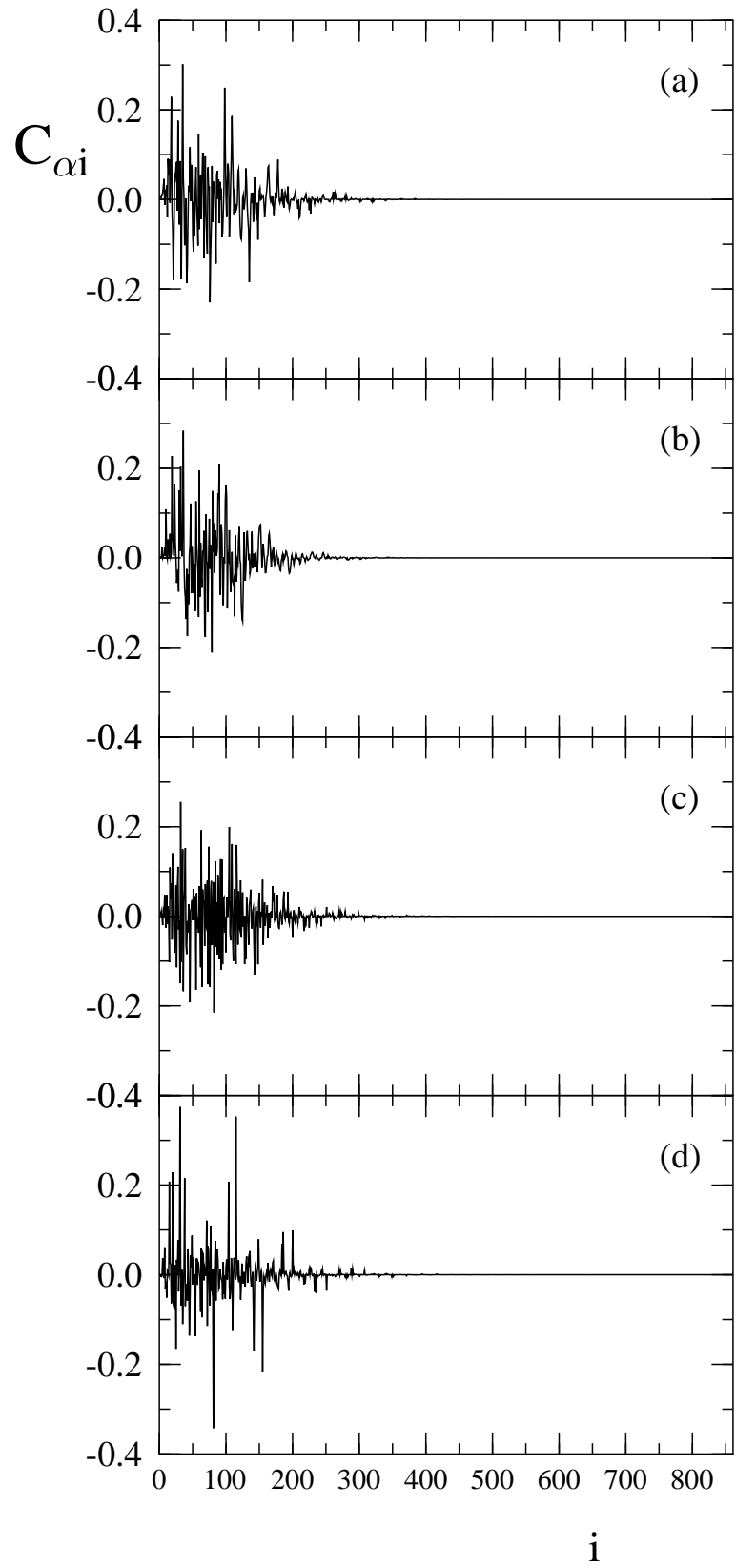
\begin{figure} \caption{ Four typical eigenstates $|\psi_{\alpha }>$ of 
the Hamiltonian $H$ for $\lambda =0.9$ in the basis $|\phi_i>$;  
(a) $\alpha$ =50, (b) $\alpha$ =51, (c) $\alpha$ =52, (d) $\alpha$ =53. } 
\end{figure}

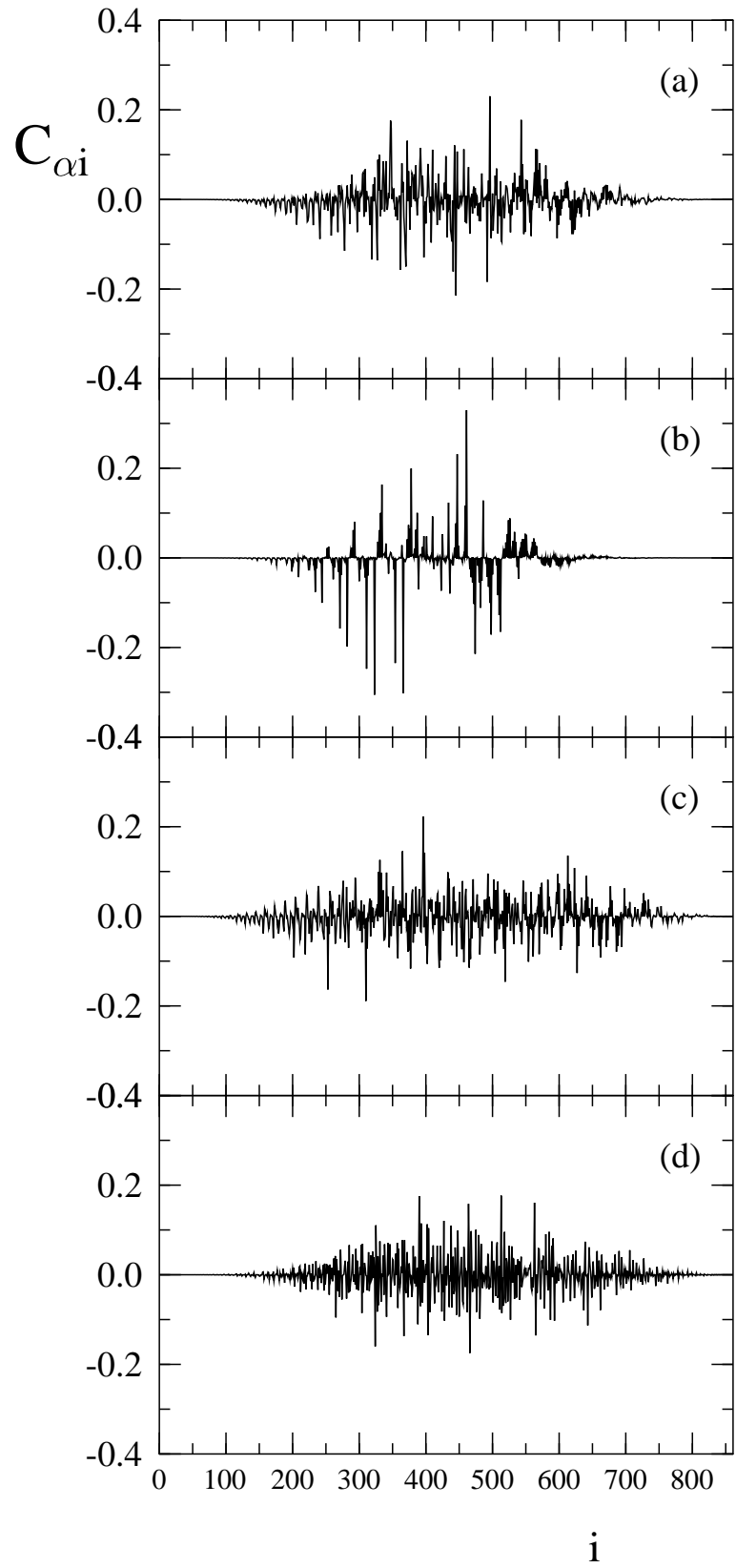
\begin{figure} \caption{ Same as in Fig.5 for 
(a) $\alpha$ = 430, (b) $\alpha$ = 431, (c) $\alpha$ = 432, (d) $\alpha$ = 433. } 
\end{figure} 

\begin{figure} \caption{ Histograms of the nearest-level-spacing 
distributions $P(s)$ for eight energy regions 
of the Hamiltonian $H$ with $\lambda =0.9$. 
(1) 1$< \alpha <$110, (2) 110$< \alpha <$220,
 (3) 220$< \alpha <$330, (4) 330$< \alpha <$440, 
(5) 440$< \alpha <$550, (6) 
550$< \alpha <$660, (7) 660$< \alpha <$770, (8) 770$< \alpha <$861. 
 The dashed and dashed-dot curves 
represent the Wigner-Dyson and Poisson distributions, respectively. 
Each histogram was obtained by diagonalizing five different 
Hamiltonians with values of $\lambda$ close to 0.9. 
} \end{figure} 

\begin{figure} \caption{ Non-zero matrix elements of the four operators 
(a) $\mu_1 K_{10}K_{10}$, (b) $\mu_2 K_{20}K_{20}$, (c) $\mu_3 K_{21}K_{20}$ 
and (d) $\mu_4 K_{12}K_{10}$. Each operator couples a 
basis state $|\phi_i>$ to at most only one  other basis state 
 $|\phi_j>$. } 
\end{figure} 

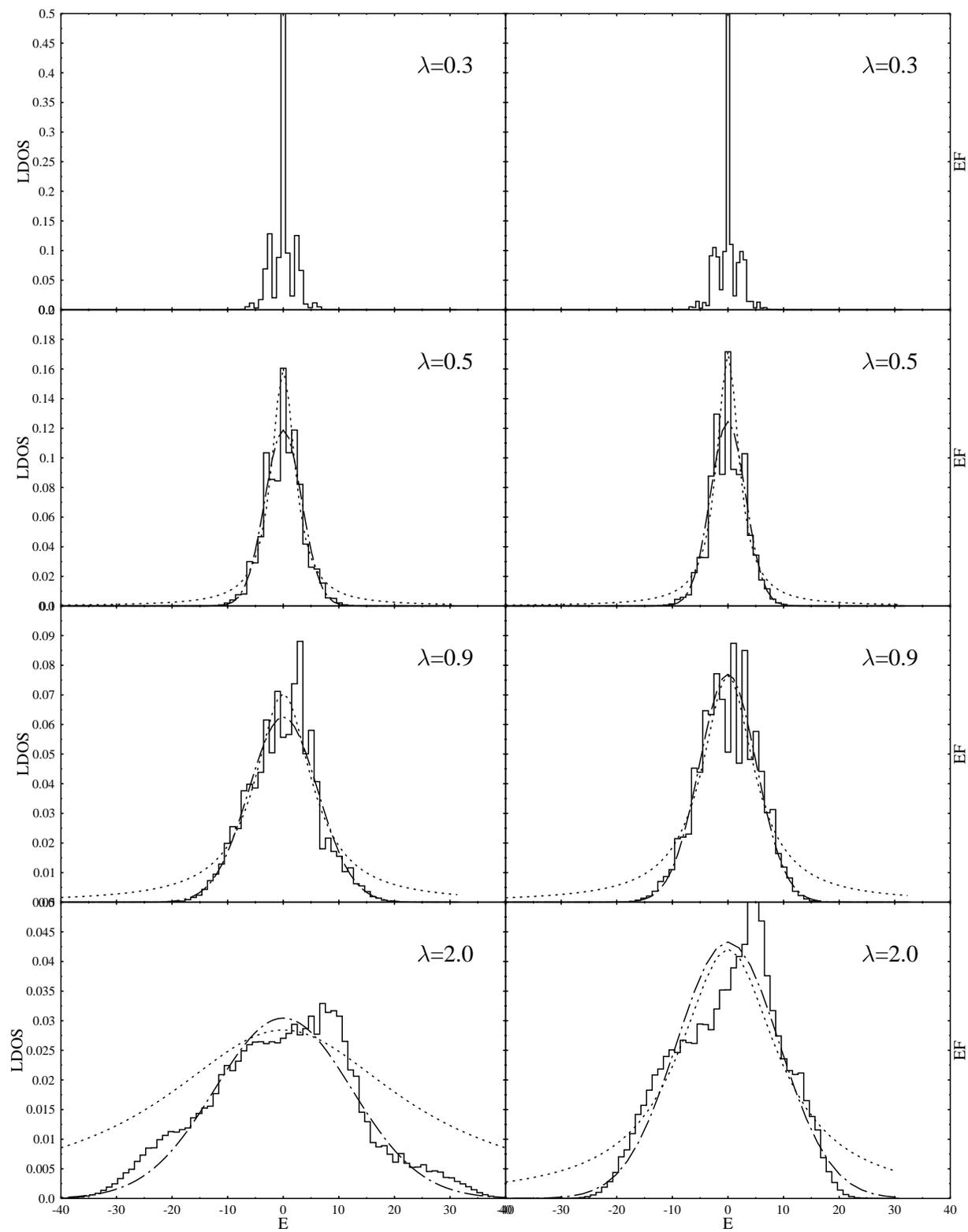
\begin{figure} \caption{ Averaged LDOS (left column histograms) and 
eigenfunctions (EF) (right column histograms) for the LMG model
with different values of $\lambda $.  
 Dashed and dashed-dot curves are fitting curves 
to the BW and Gaussian forms, respectively. 
} \end{figure} 

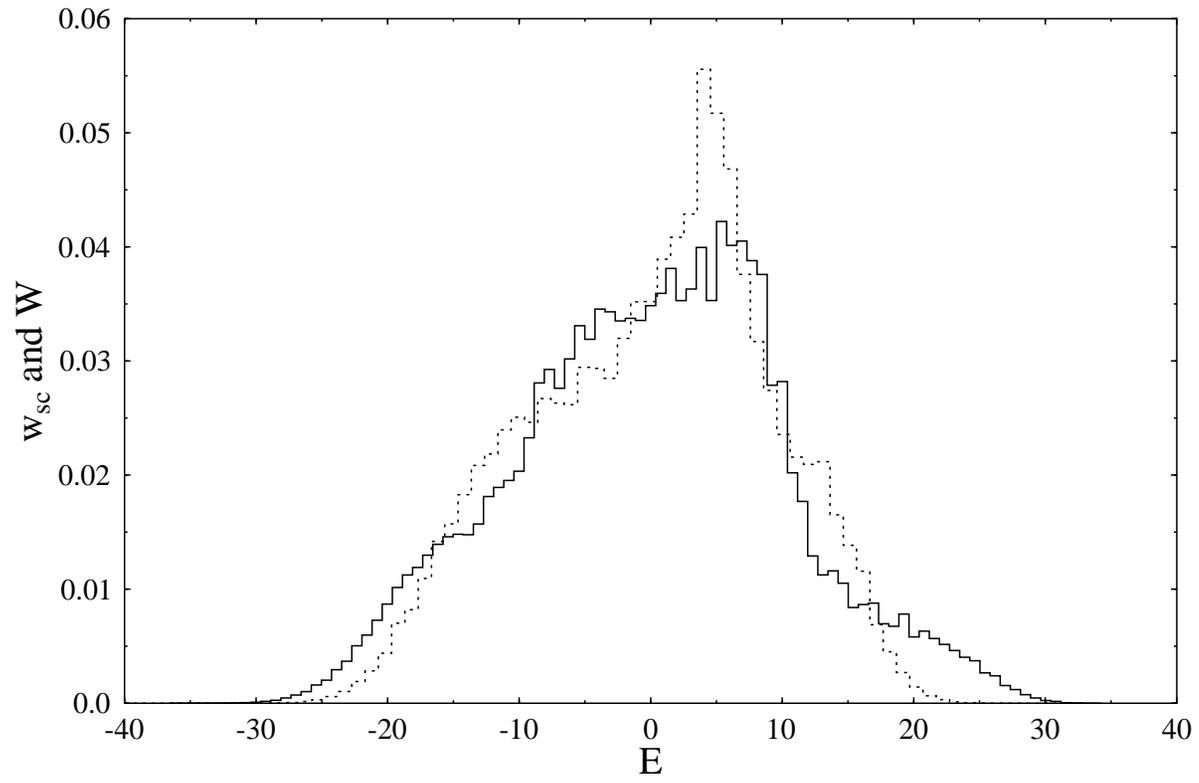
\begin{figure} \caption{ A comparison between the rescaled 
 LDOS $w_{sc}(E)$ (solid histogram) 
and the EF $W(E^0)$ for $\lambda =2$ 
(dashed histogram). } 
\end{figure} 

\begin{figure} \caption{ Classical counterparts of the LDOS (dashed-dot  
curves) for (a) $\lambda =0.3$, (b) $\lambda =0.5$, (c) $\lambda =0.9$ 
and (d) $\lambda =2$. For comparison the corresponding LDOS are also 
plotted (solid histograms). } 
\end{figure} 

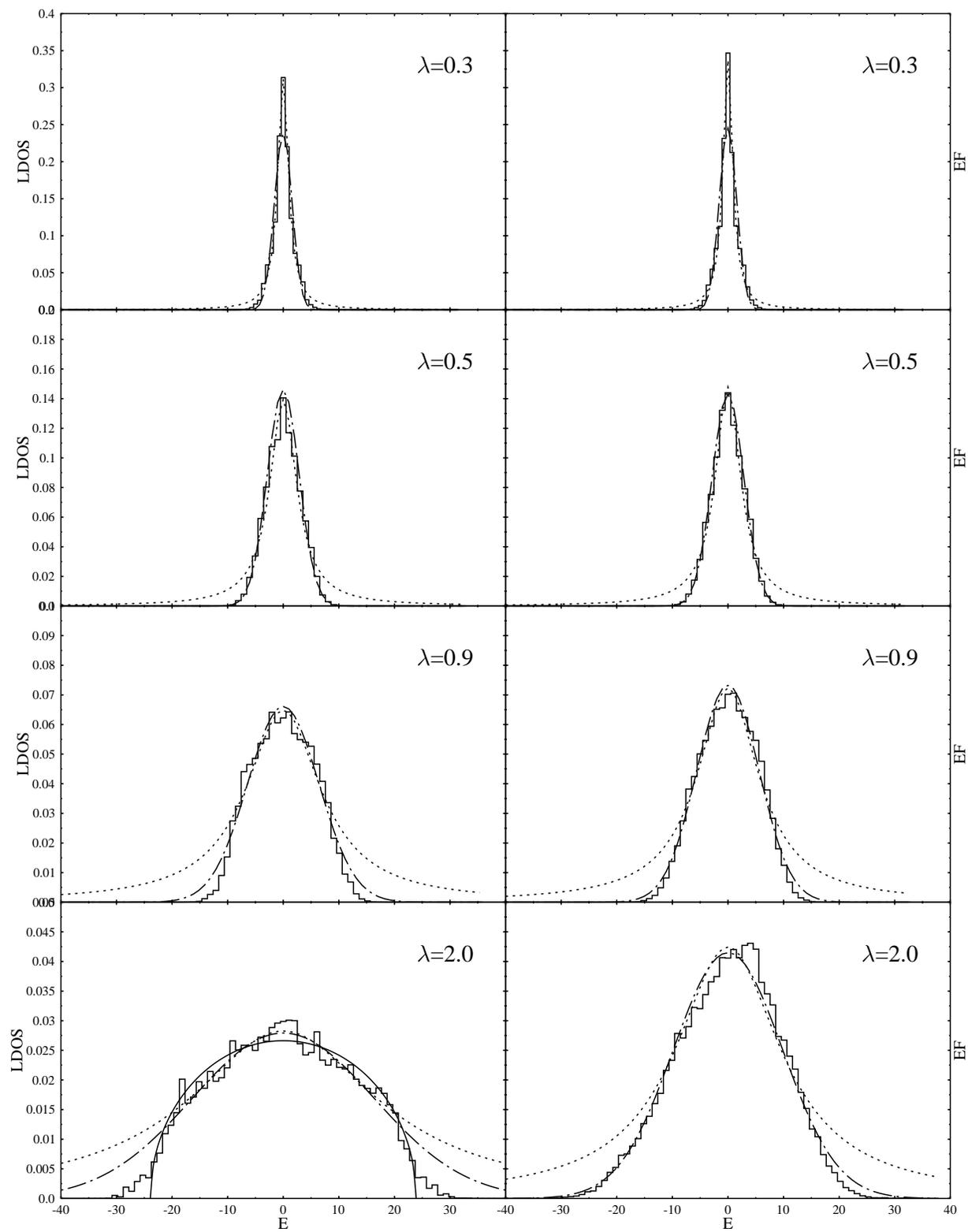
\begin{figure} \caption{ Same as in Fig.9 but for the band random matrix 
model $H_{ran}$. The solid curve for the  LDOS at $\lambda =2.0$ is the 
fitting curve to the semicircle law Eq. (27) with $ R_0 \approx 
23.9. $ }
\end{figure} 

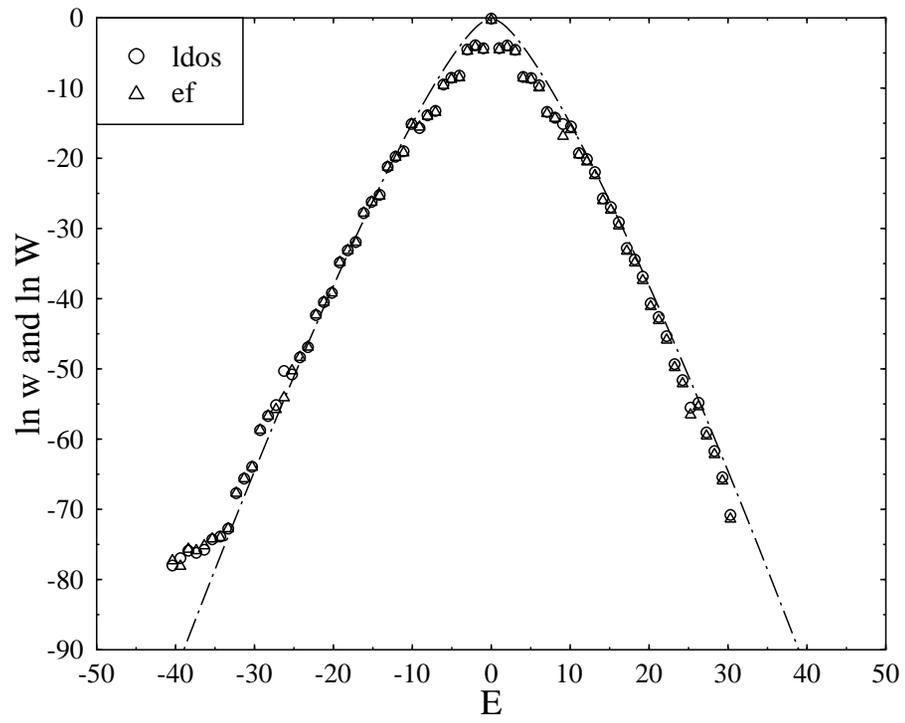
\begin{figure} \caption{ Numerically computed LDOS (circles) and EF
 (triangles) for $\lambda =0.1$. Each value represents an 
average over 200 states.  
The dashed-dot curve gives the analytical prediction (30) 
(with the appropriate normalization factor) for the long tails.}  
\end{figure} 

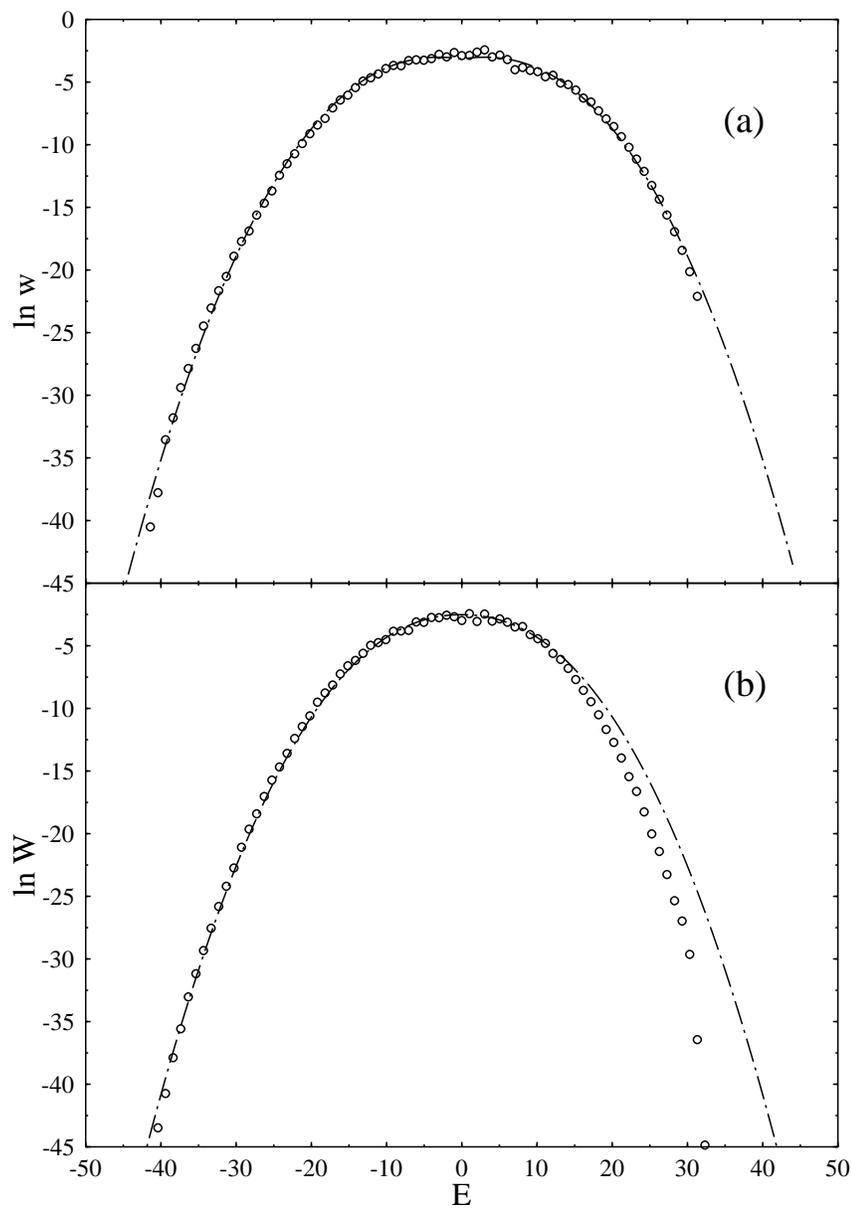
\begin{figure} \caption{ 
(a) LDOS (circles) for $\lambda =0.9$ and the fitting curve (dashed-dot 
curve) given by  
 Eqs.\ (30) 
and \ (33) with  $ x_0=-13$. 
(b) Similar to (a) for the EF (circles) with $x_0=-16$. 
} \end{figure} 

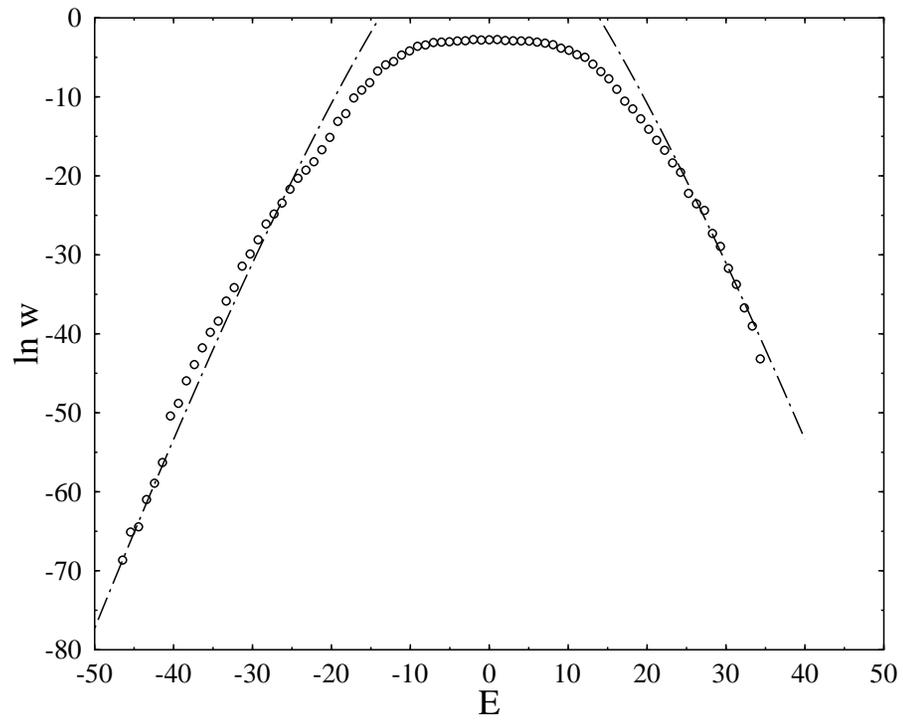
\begin{figure} \caption{ LDOS (circles) of $H_{ran}$ in Eq.\ (26) 
for $\lambda =0.9$ and the theoretical 
 prediction \ (30) (dashed-dot curve) 
 for its long tails. 
} \end{figure}

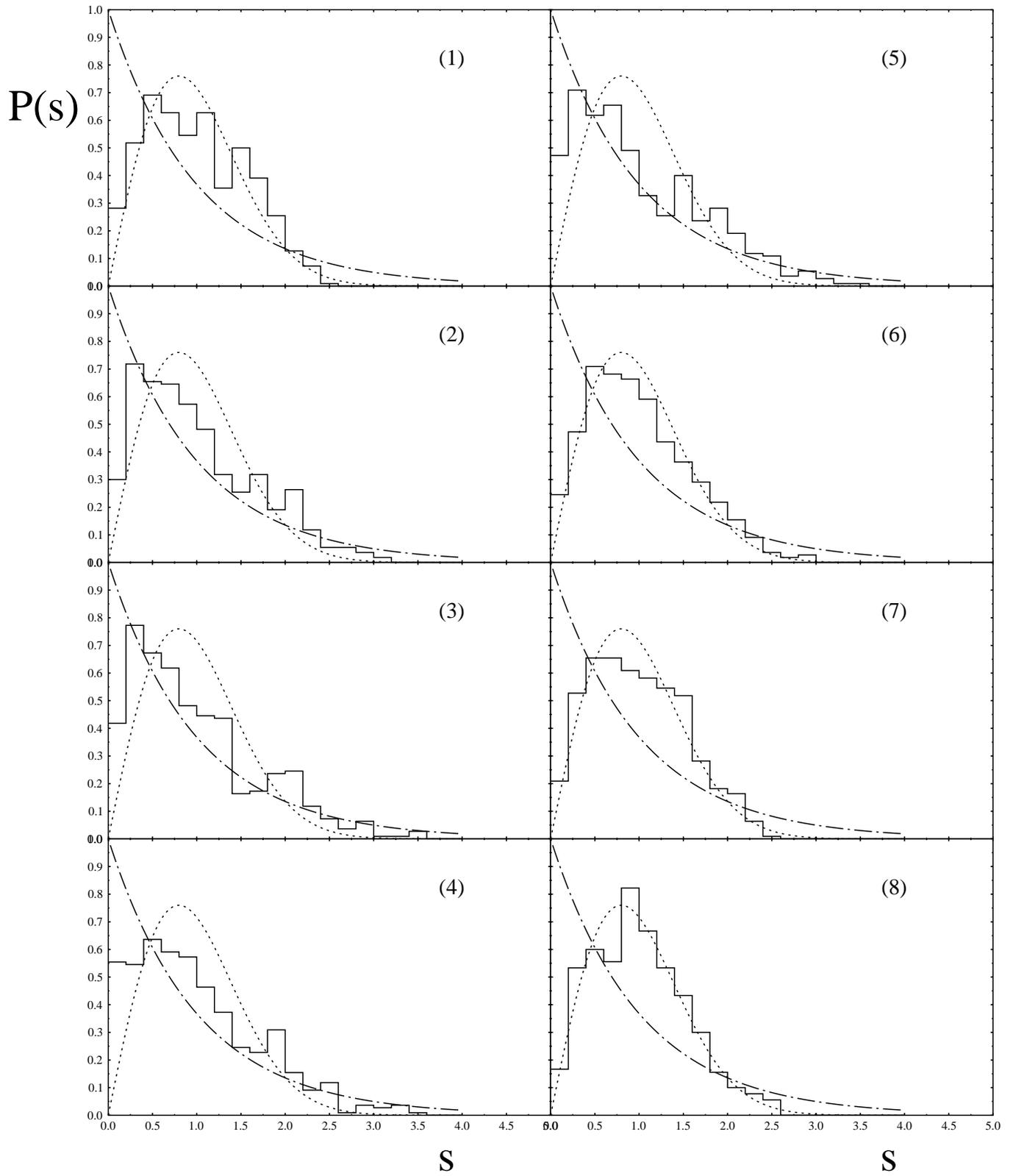
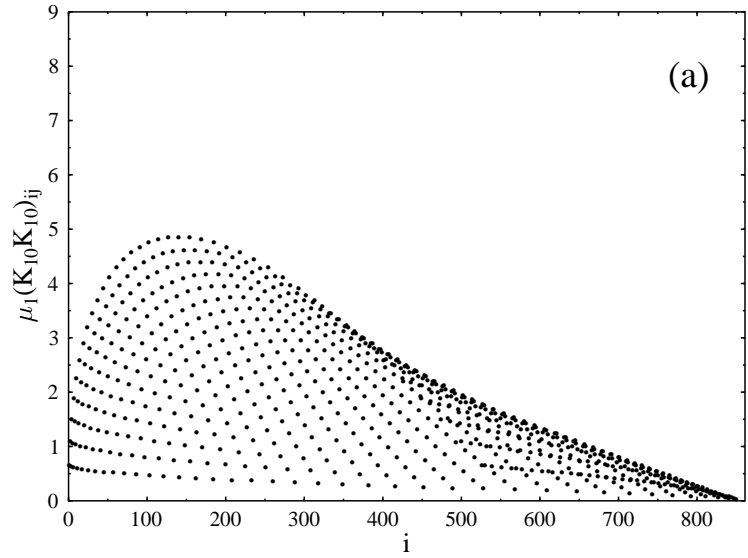
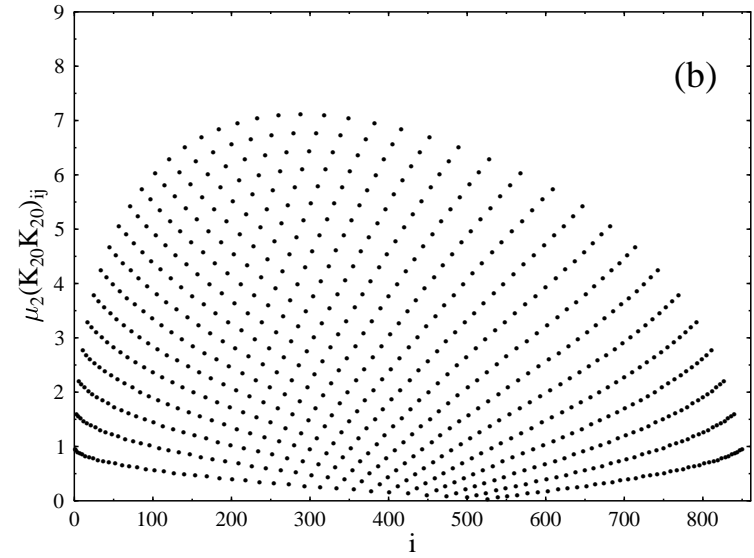
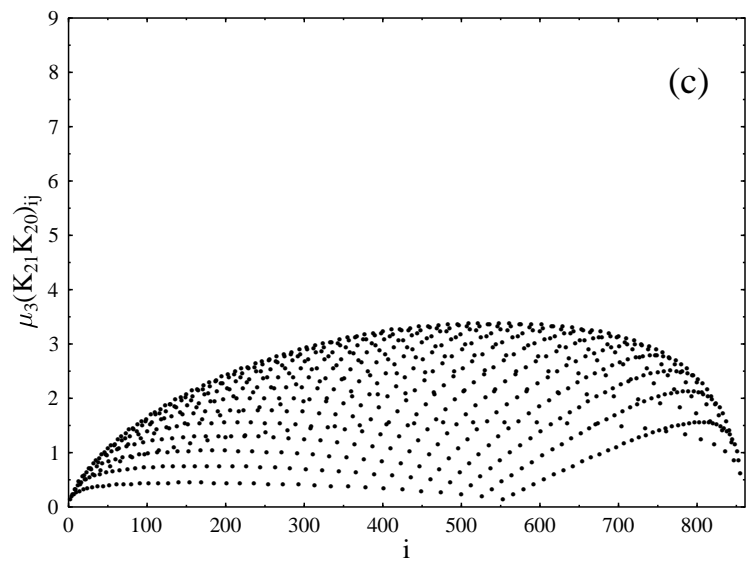
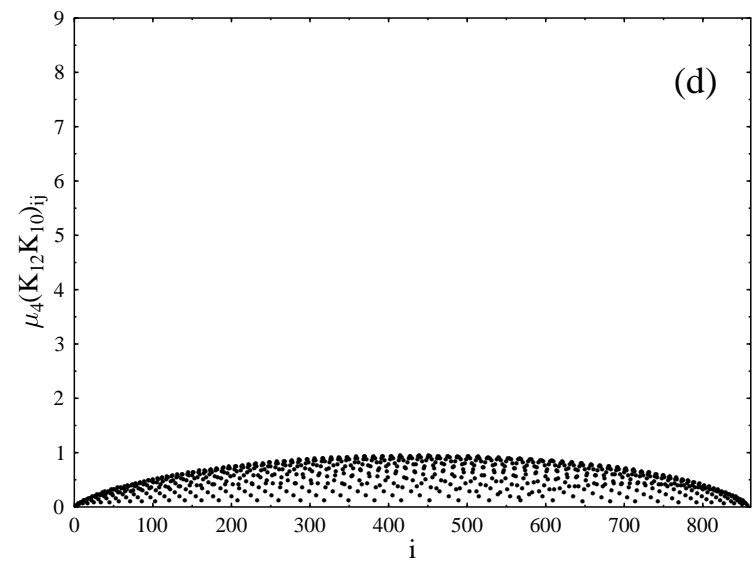
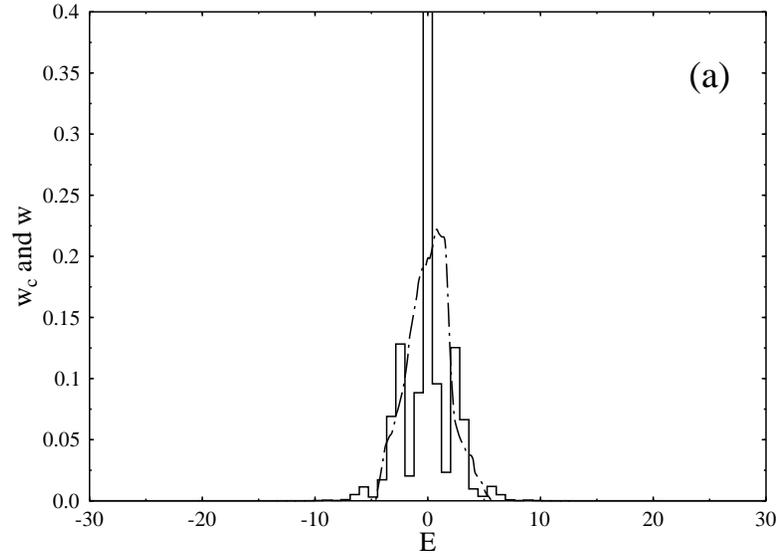
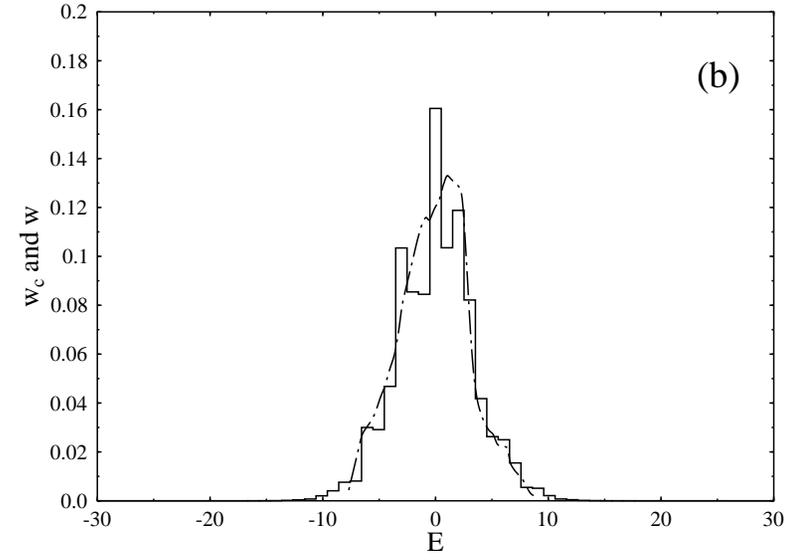
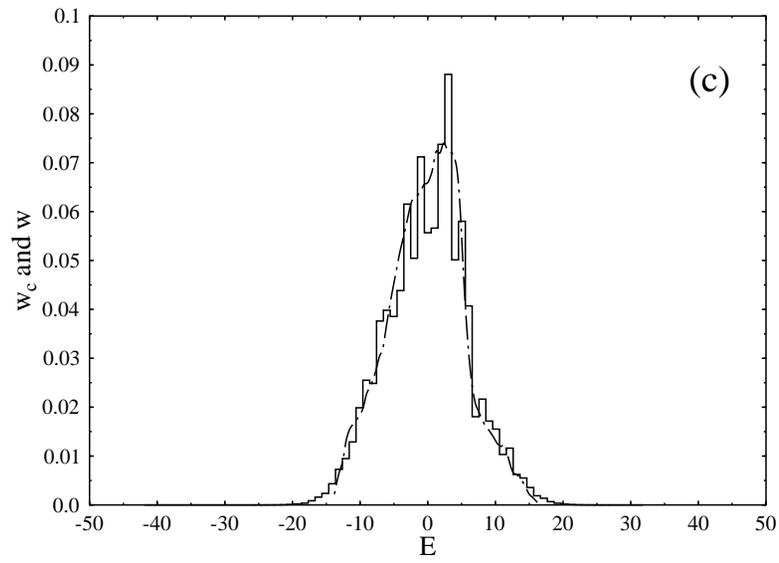
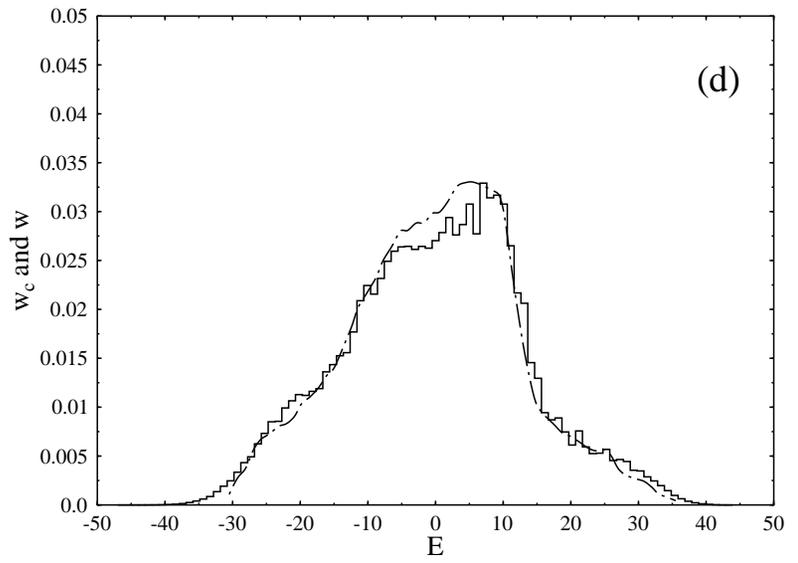
\end{document}